\documentclass[aps,prb,twocolumn,draft,groupedaddress,showpacs]{revtex4}

\usepackage{amssymb}
\usepackage{amsfonts}
\usepackage{amsmath}
\usepackage{epsf}

\begin{document}

\title{Weak localization and conductance fluctuations in a quantum dot with
parallel magnetic field and spin-orbit scattering}
\author{Jan-Hein Cremers}
\affiliation{Lyman Laboratory of Physics, Harvard University, 
Cambridge MA 02138}

\author{Piet W.\ Brouwer}
\affiliation{Laboratory of Atomic and Solid State Physics, Cornell 
University, Ithaca, NY 14853-2501}

\author{Vladimir I.\ Fal'ko}
\affiliation{Physics Department, Lancaster University, Lancaster L41
  4YB, United Kingdom}

\date{\today}

\begin{abstract}
In the presence of both spin-orbit scattering and a magnetic field the
conductance of a chaotic GaAs quantum dot displays quite a rich behavior.
Using a Hamiltonian derived by Aleiner and Fal'ko [Phys. Rev. Lett. 
\textbf{87}, 256801 (2001)] we calculate the weak localization
correction and the covariance of the conductance, as a function of
parallel and perpendicular magnetic field and spin-orbit coupling
strength. We also show how the combination of an in-plane 
magnetic field and spin-orbit scattering gives rise to a component
to the magnetoconductance that is anti-symmetric with respect to 
reversal of the perpendicular component of the magnetic field and
how spin-orbit scattering leads to a ``magnetic-field echo'' in the
conductance autocorrelation function. Our results can be used for
a measurement of the Dresselhaus and Bychkov-Rashba spin-orbit 
scattering lengths in a GaAs/GaAlAs heterostructure.

\end{abstract}

\maketitle

\section{Introduction}

A two-dimensional (2D) electron gas
offers ample opportunity for manipulation of the electron's spin and orbital
state. In the absence of spin-orbit coupling, such manipulation can
be performed separately on the electron spin and orbital degree of freedom,
by means of magnetic fields of various orientations and electrostatic gates.
On one hand, a magnetic field parallel to the plane of the 2D gas lifts the
spin degeneracy and allows for a measurement that distinguishes between the
transport properties of \textquotedblleft up\textquotedblright\ and
\textquotedblleft down\textquotedblright\ spins: As long as the confining
potential that determines the 2D electron gas is sharp, it does not
significantly affect the electron's orbital degrees of freedom. On the other
hand, gate voltages and a weak magnetic field perpendicular to the quantum
well act on the orbital degrees of freedom and do not couple significantly
to the electron's spin.\cite{mesoreview1}

Spin-orbit scattering couples the spin and orbital degrees of freedom.
Conventionally, the main signature of spin-orbit coupling in the 2D
electron gas in GaAs heterostructures is the observation of weak
anti-localization, a small positive quantum correction to the conductivity
that is suppressed by a magnetic field.\cite{Hikami} Weak anti-localization
is the counterpart of weak localization, a negative correction to
the conductivity at zero magnetic field caused by the constructive
interference of back-scattered electron waves in a phase-coherent disordered
conductor in the absence of spin-orbit scattering.\cite{Bergmann} Only
recently, spin-dependent phase coherent transport was considered for a
quantum dot, a small island of the electron gas confined by gates 
\cite{Khaetskii}. Motivated by an experiment by Folk 
\emph{et al.},\cite{Folk}
theoretical works by Halperin \emph{et al.}\cite{Halperin} and by two of the
authors,\cite{AF} have shown that the combination of spin-orbit scattering
and a spin degeneracy lifting parallel magnetic field leads to a remarkably
rich structure of quantum interference phenomena in III-V semiconductor 
quantum dots, that far surpasses a simple
interpolation between weak localization and weak antilocalization physics.

Not only does the crossover regime between weak localization and
anti-localization represent an interesting issue of its own,
understanding features imposed upon quantum transport by spin-orbit
coupling is needed for a correct extraction of the decoherence time
from experimental data. For electrons in a small dot with weak but
finite spin-orbit coupling, the the crossover of weak localization to
anti-localization manifests itself as a suppression of the
zero-temperature value of localization correction to the dot
conductance. This suppression of the weak localization correction may
be misinterpreted as a saturation of the decoherence time
$\tau_{\phi}$ at low temperatures. Without a priori knowledge of the
spin-orbit coupling constants for a given semiconductor dot, one must
use the features of spin-orbit coupling induced interference effects
in order to identify the origin of what can be mistaken for a suppression
of the interference part of conductance, when the observed weak
localization correction is less than the
prediction of the quantum transport theory for to the crossover
between orthogonal and unitary symmetry classes.

In this publication, we present a detailed quantitative analysis of
the influence of the coupling between electron spin and orbital motion
in a two-dimensional semiconductor on the interference corrections to
the conductance of a quantum dot. We consider the effects of
enhanced/suppressed back-scattering and the variance and correlation
properties of universal conductance fluctuations. The calculations,
which are performed in the framework of the random scattering matrix
approach of random matrix theory, are described in Sections
\ref{sec:2} and
\ref{sec:3}. In Section \ref{sec:2}, 
we present the analysis of the average conductance
of a chaotic dot, Section \ref{sec:3} is devoted to conductance
fluctuations. For technical reasons, we limit our attention to the
case of a large number of channels, $N_{1}$ and $N_{2}$ in the point
contact connecting the dot to the bulk, though all qualitative
features we discover for $N\gg 1$ would persist in the case of $N\sim
1$. These two technical sections are preceded by a qualitative
discussion (Sec.\ \ref{sec:1}) and followed by an analysis of the 
effect of a
spatially non-uniform spin-orbit coupling strength (Sec.\ 
\ref{sec:nonuniform}).

\section{Interplay between SO coupling and Zeeman splitting in quantum
  dots} \label{sec:1}

In GaAs/AlGaAs heterostructures, spin-orbit coupling owes its
existence to the asymmetry of the potential confining the 2DEG and the lack
of inversion symmetry in the crystal structure, leading to the
Bychkov-Rashba and Dresselhaus terms in the Hamiltonian. In the theory
we present below, we consider electrons in a heterostructure or quantum well
lying in the (001) crystallographic plane of a zinc-blend type semiconductor
and choose coordinates $x_{1},x_{2}$ along crystallographic directions 
$\mathbf{\hat{e}}_{1}=[110]$ and $\mathbf{\hat{e}}_{2}=[1\bar{1}0]$
(coordinate $x_{3}$ is perpendicular to the plane of the two-dimensional
electron gas). The effective two-dimensional Hamiltonian of
electrons in a quantum dot takes the form 
\begin{eqnarray}
  \mathcal{H} &=&\frac{1}{2m}\left[ \left(
  p_{1}-eA_{1}-\frac{\sigma_{2}}{2\lambda_{1}}\right)^{2}+
  \left( p_{2}-eA_{2}+\frac{\sigma_{1}}{2\lambda_{2}}\right)^{2}
  \right]  \notag \\
  && \mbox{}+V(\mathbf{r})+\frac{1}{2}\mu_{B}g \mathbf{B} 
  \cdot \mbox{\boldmath $\sigma$}. \label{eq:FullHam}
\end{eqnarray}%
where $\mathbf{B}$ is the magnetic field, 
$\mathbf{A}=B_{3}(\mathbf{\hat{e}}_{3}\times \mathbf{r})/2c$ 
is the vector potential corresponding to the
component of the magnetic field perpendicular to the plane of the
two-dimensional electron gas, and $\mbox{\boldmath $\sigma$}=(\sigma_{1},\sigma
_{2},\sigma_{3})$ the vector of Pauli matrices. The potential $V$ both
confines the electrons to the quantum dot and describes elastic scattering
from non-magnetic impurities in the dot. The two length scales, 
$\lambda_{1}$ and $\lambda_{2}$ are associated with
spin-orbit coupling for an electron moving along the principal
crystallographic directions $\mathbf{\hat{e}}_{1}$ and 
$\mathbf{\hat{e}}_{2}$,\cite{RashbaDresselhaus} and characterize
the length at which spin of an initially polarized electron would
precess with an angle $2\pi $.

For a dot with homogeneous electron density and, therefore, parameters of
confining potential, spin-orbit coupling parameters $\lambda_{1}$ and 
$\lambda_{2}$ are independent of the position. In that case, the spin-orbit
coupling takes the form of a spin-dependent \textquotedblright vector
potential\textquotedblright , which is non-abelian since the Pauli matrices
do not commute with each 
other.\cite{Meir,MathurStone,Oreg} If $\lambda_{1}$ and 
$\lambda_{2}$ are large compared to the dot size $L$, the non-commutativity
of this spin-dependent vector potential involves higher powers of $L/\lambda
_{1,2}$, so that the spin-orbit coupling can be \textquotedblleft gauged
out\textquotedblright\ to leading order in $L/\lambda_{1,2}$. This
may be achieved through the unitary transformation $\psi (\mathbf{r})=
U(\mathbf{r})\tilde{\psi}(\mathbf{r})$, 
\begin{eqnarray}
  U &=&\exp \left( \frac{ix_{1}\sigma_{2}}{2\lambda_{1}}-\frac{ix_{2}\sigma
_{1}}{2\lambda_{2}}\right) 
  \nonumber \\ &=& \cos R+\left( \frac{ix_{1}\sigma_{2}}{2\lambda
_{1}}-\frac{ix_{2}\sigma_{1}}{2\lambda_{2}}\right) \dfrac{\sin R}{R},
\label{eq:U} 
\\
  R &=&\sqrt{\left( \frac{x_{1}}{2\lambda_{1}}\right) ^{2}+
  \left( \frac{x_{2}}{2\lambda_{2}}\right) ^{2}},
\end{eqnarray}%
which performs a position-dependent rotation of the spin out of the plane of
the two-dimensional electron gas to the locally adjusted
frame.\cite{AF} 
Below, we assume that $R\ll 1$, and use $R$ as a small
parameter to derive the form of the Hamiltonian $\tilde{\mathcal{H}}
=U^{\dagger }\mathcal{H}U$ in the locally rotated spin-frame, 
\begin{eqnarray}
\tilde{\mathcal{H}}&=&\frac{1}{2m}\left( -i\hbar \nabla \mathbf{-}e\mathbf{A}-
\mathbf{a}_{\bot }-\mathbf{a}_{\Vert }\right) ^{2}+h^{\mathrm{Z}}+h_{\bot }^{
\mathrm{Z}}+V(\mathbf{r}).
  \nonumber \\
  \label{eq:Hamtilde} \label{eq:hamiltonian}
\end{eqnarray}%
Here we introduced the spin-dependent vector potential
$$
\mathbf{a}_{\bot }=\frac{\sigma_{3}}{4\lambda_{1}\lambda_{2}}[\mathbf{
\hat{e}}_{3}\times \mathbf{r}]
$$
that has the same form as a magnetic field $\pm e c \mathbf{\hat{e}_3}
/2 \lambda_{1}\lambda_{2}$ with opposite directions for electrons with
spins ``up'' and ``down'' in the new
spin-frame, and we also abbreviated
\begin{eqnarray}
  \mathbf{a}_{\Vert } &=&\frac{1}{6\lambda_{1}\lambda_{2}}
  \left( \frac{x_{1}\sigma_{1}}{\lambda_{1}}
  + \frac{x_{2}\sigma_{2}}{\lambda_{2}}\right)
  [\mathbf{\hat{e}}_{3}\times \mathbf{r}]  \notag \\
  h^{\mathrm{Z}} &=&\frac{1}{2}\mu_{B}g\mathbf{B}\cdot 
  \mbox{\boldmath $\sigma$}, 
  \notag \\
  h_{\bot }^{\mathrm{Z}} &=&-\mu_{B}g
  \left( \frac{B_{1}x_{1}}{2\lambda_{1}}
  + \frac{B_{2}x_{2}}{2\lambda_{2}}\right) \sigma_{3}.
\end{eqnarray}%

The Hamiltonian (\ref{eq:Hamtilde})
describes electrons in a rotated spin frame. It
contains all relevant terms that lift the high degree of symmetry of a
system with uncoupled orbital and spin degrees of freedom, to leading
order in the small parameter $L/\lambda_{1,2}$. In these equations,
we omitted sub-leading terms of higher order in $L/\lambda_{1,2}$
that do not affect the symmetry of the Hamiltonian.

As long as the rate at which electrons escape from the quantum dot into the
leads is much smaller than the Thouless energy $E_{\mathrm{Th}}$, transport
properties of the quantum dot can be calculated using random matrix 
theory.\cite{Beenakker97}
 The use of random matrix theory requires an
analysis of the symmetries of the scattering matrix, which are set by the
relative magnitudes of characteristic energy scales for each of the terms in
the Hamiltonian (\ref{eq:hamiltonian}) and the escape rate. These energy
scales are 
\begin{eqnarray}
\varepsilon_{B} &=&\kappa E_{\mathrm{Th}}\left( {eB_{3}L^{2}}/2{\hbar }
\right) ^{2},  \notag \\
\varepsilon_{\bot }^{\mathrm{so}} &=&\kappa E_{\mathrm{Th}}\left(
	    {L^{2}}/{4\lambda ^{2}}\right) ^{2},  \notag \\
\varepsilon_{\Vert }^{\mathrm{so}} &=&\kappa ^{\prime }\left( 
{L^{2}/4\lambda ^{2}}\right) \varepsilon_{\bot }^{\mathrm{so}},
  \label{eq:E_so} \label{eq:kappa}  \\
\varepsilon ^{\mathrm{Z}} &=&\mu_{B}gB,  \notag \\
\varepsilon_{\bot }^{\mathrm{Z}} &=&\frac{\kappa^{\prime \prime}
(\varepsilon^{\mathrm{Z}})^{2}}{E_{\mathrm{Th}}}\left( L^{2}/4\lambda
^{2}\right) ,  \notag
\end{eqnarray}%
for the orbital contribution of the magnetic field, the spin-orbit
terms $a_{\Vert }$ and $a_{\bot }$, and the Zeeman coupling terms 
$h^{\mathrm{Z}}$ and $h_{\bot }^{\mathrm{Z}}$, respectively. 
For weak uniform spin-orbit coupling in a small dot such that 
$L/\lambda_{1,2}\ll 1$, one has the strong inequalities 
\begin{equation}
\varepsilon_{\Vert }^{\mathrm{so}}\ll \varepsilon_{\bot }^{\mathrm{so}},\
\ \varepsilon_{\bot }^{\mathrm{Z}}\ll \varepsilon ^{\mathrm{Z}}.
\label{eq:ineq}
\end{equation}
Here $E_{\rm Th}$ is the Thouless energy, which is the largest
energy scale in the problem, $\lambda ^{2}=\lambda
_{1}\lambda_{2}$, and $\kappa $, $\kappa ^{\prime }$, and $\kappa ^{\prime
\prime }$ are coefficients of order unity. The coefficients $\kappa ^{\prime
}$ and $\kappa ^{\prime \prime }$ depend on the sample geometry and on the
ratio $\lambda_{1}/\lambda_{2}$ of the spin-orbit lengths. The coefficient 
$\kappa ^{\prime \prime }$ also depends on the direction of the magnetic
field. The coefficient $\kappa $ depends on the sample geometry only; it
appears both for the orbital contribution of the magnetic field and for the
spin-orbit term $a_{\bot }$ because both terms have the same spatial
dependence. For a circular quantum dot with radius $L$, mean free path 
$l\ll L$, Fermi velocity $v_{F}$, and $\lambda_{1}=\lambda_{2}=\lambda $,
one has $E_{\mathrm{Th}}=\hbar v_{F}l/2L^{2}$, $\kappa =2$, $\kappa ^{\prime
}=1/3$ and $\kappa ^{\prime \prime }\approx 0.292$, see App.\ \ref{app:1}.

If the spin-orbit coupling is non-uniform, i.e., when the spin-orbit lengths 
$\lambda_{1}$ and $\lambda_{2}$ depend on position, the \textquotedblleft
spin-dependent vector potential\textquotedblright\ in Eq.\
(\ref{eq:FullHam}) can no longer be gauged away to leading order 
in $L/\lambda_{1,2}$. A
spatially non-uniform spin-orbit coupling can be created intentionally, with
the help of a metal gate parallel to the two-dimensional electron gas that
changes the asymmetry of the quantum well.\cite{Shayegan,Marcus,Zumbuehl},
or arise as a by-product of the confining gates. Generically, the
non-uniformity gives rise to a term in the Hamiltonian of the same symmetry
as $\mathbf{a}_{\Vert }$ in Eq.\ (\ref{eq:hamiltonian}).\cite{BCH} Hence,
the main effect of non-uniformities is to increase the corresponding energy
scale $\varepsilon_{\Vert }^{\mathrm{so}}$ by an amount of order 
$E_{\mathrm{Th}}(L/\lambda_{\mathrm{fl}})^{2}$, where 
$1/\lambda_{\mathrm{f}}$
is a measure of the fluctuations of the spin-orbit coupling $\lambda
_{1,2}^{-1}$.\cite{BCH} For very small quantum dots, this increase of 
$\varepsilon_{\Vert }^{\mathrm{so}}$ may eventually reverse the first
inequality in Eq.\ (\ref{eq:ineq}). 

It is the existence of two energy scales each to describe the strength of
the spin-orbit and Zeeman terms in the Hamiltonian that leads to the rich
parameter dependence of the conductance distribution of a chaotic quantum
dot. When all relevant energy scales are either much larger or much smaller
than the escape rate, the scattering matrix has a well defined symmetry, and
the conductance distribution can be found using symmetry considerations
alone.\cite{AF} 
It is the goal of this paper to address the general parameter regime, where
an interpolation between the various symmetry classes is called for. Our
results are important for a quantitative analysis of recent experiments by
Zumb\"{u}hl \emph{et al.},\cite{Zumbuehl} and for the development of
techniques for experimental determination of spin-orbit coupling 
parameters. Looking
into various regimes, we show how to extract such information from (a) weak
localization measurements in the presence of an in-plane magnetic field; (b)
an analysis of how the in-plane magnetic field gives rise to a
component of the magnetoconductance that is anti-symmetric
in the perpendicular magnetic field; and (c) from the possible 
observation of a \textquotedblleft
magnetic field echo\textquotedblright\ in the auto-correlation function of
conductance fluctuations.

The latter prediction is made on the basis of the following
theoretical observation. After the unitary transformation (\ref{eq:U}), the
leading spin-orbit term $\mathbf{a}_{\bot }^{\mathrm{so}}$ has precisely the
same spatial dependence as the vector potential due to the orbital
magnetic field (in a symmetric gauge). Hence, 
$\mathbf{a}_{\bot }^{\mathrm{so}}$ can be regarded as an effective 
magnetic field perpendicular
to the plane of the quantum well and of opposite sign for the two spin
directions. A suitably chosen orbital magnetic field can balance this
effective field, or change its sign for one spin direction, leading to a
partial re-appearance of weak localization and to a \textquotedblleft
magnetic field echo\textquotedblright in the auto-correlation
function of conductance fluctuations. The magnetic field echo
appears in the magnetoconductance autocorrelation function for
a magnetic field difference
\begin{equation}
  \Delta B_3 = \frac{\hbar}{e \lambda_1 \lambda_2};
\end{equation}
The shift of weak localization peaks is by half this field
strength.

According to the Onsager's relations, the two-terminal conductance of a
non-magnetic system is symmetric with respect to reversal of the
magnetic field $\mathbf{B}$. In the absence of spin-orbit coupling, 
the symmetry with respect to the reversal of the perpendicular
magnetic field component $B_{3}$ would persist even in the presence 
of Zeeman splitting caused in a 2D
electron gas by the in-plane magnetic field $\mathbf{B}_{\Vert}$
(excluding effects arising from the finite width of the quantum
well.\cite{FJ}) However,
in the presence of even a weak SO coupling, the exact symmetry exists
only if both $B_{3}$ and $B_{\Vert }$ are reversed simultaneously, 
whereas the reversal of the perpendicular component only generates an
antisymmetric contribution to the magnetoconductance with variance   
\begin{equation}
\mathrm{var}\left[ G(B_{3},B_{\Vert })-G(-B_{3},B_{\Vert })\right] \sim 
\dfrac{\left( \varepsilon ^{\mathrm{Z}}\right) ^{2}
\varepsilon_{\Vert }^{\mathrm{SO}}}{\left( N\Delta /2\pi \right)^{3}} 
\left( \dfrac{e^{2}}{h}\right) ^{2},  
  \label{eq:AsymmetricPart}
\end{equation}%
where $N$ is the number of channels in the leads, $\Delta $ is the
mean level spacing in the dot, and $\langle G \rangle$ is the
ensemble-averaged conductance of the quantum dot. Since the
fluctuating antisymmetric contribution the conductance,
$G(B_{3},B_{\Vert })- G(-B_{z},B_{\Vert })$ is linear in the in-plane
magnetic field, $\varepsilon ^{\mathrm{Z}}\approx g\mu_{B}B_{\Vert
}$, it dominates over the conductance asymmetry caused by
inter-subband mixing due to the in-plane magnetic field for
intermediate magnetic fields strengths,\cite{FJ} which is proportional
to $B_{\Vert }^{3}$.

\section{Average conductance}

\label{sec:2} 

The starting point of our calculation is the Landauer formula for the
two-terminal conductance $G$ of the quantum dot at zero temperature, 
\begin{equation}
G = {\frac{2e^{2}}{h}}{\frac{N_{1}N_{2}}{N}} - {\frac{e^{2}}{h}}
\mbox{tr}\,S\Lambda S^{\dagger }\Lambda,  \label{eq:Landauer}
\end{equation}
where the diagonal matrix $\Lambda $ has elements 
\begin{equation}
\Lambda_{jj}=\left\{ 
\begin{array}{ll}
N_{2}/N, & j=1,\ldots ,N_{1}, \\ 
-N_{1}/N, & j=N_{1}+1,\ldots ,N.
\end{array}
\right.
\end{equation}
As a matrix of complex numbers, the scattering matrix $S$ has dimension $2N
= 2(N_1 + N_2)$; it may also be seen as an $N \times N$ matrix of
quaternions, which are $2 \times 2$ matrices with special rules for complex
conjugation and transposition.\cite{Mehta}

\begin{figure}[t]
\epsfxsize=0.9\hsize 
\epsffile{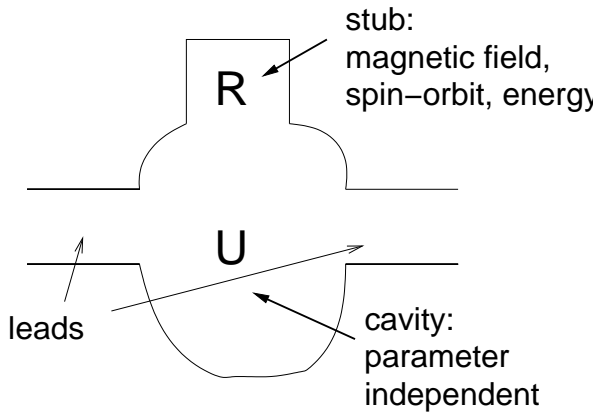} 
\caption{Illustration of the quantum dot with stub. Energy dependent
scattering from the quantum dot in the presence of spin-orbit coupling and a
magnetic field is modeled as scattering from a chaotic quantum dot with a
parameter independent scattering matrix $U$, with a stub described by a
reflection matrix $R$ which depends on energy, spin-orbit coupling and the
magnetic field. }
\label{fig:1a}
\end{figure}

In order to calculate the average of the conductance at Fermi energy 
$\varepsilon $ and magnetic field $\mathbf{B}$, or the covariance at energies 
$\varepsilon $ and $\varepsilon ^{\prime }$ and magnetic fields $\mathbf{B}$
and $\mathbf{B}^{\prime }$, it is sufficient to compute the average 
\begin{equation}
\langle S_{kl;\mu \nu }(\varepsilon ,\mathbf{B})S_{k^{\prime }l^{\prime
};\mu ^{\prime }\nu ^{\prime }}(\varepsilon ^{\prime },\mathbf{B}^{\prime
})^{\ast }\rangle ,  \label{eq:corr}
\end{equation}%
where roman indices refer to the propagating channels in the leads, and
Greek indices refer to spin. Hereto, we need a statistical description of
the scattering matrix $S$ in the presence of spin-orbit scattering and for
arbitrary values of the magnetic field $\mathbf{B}$ and Fermi energy 
$\varepsilon $. Within the \textquotedblleft random scattering matrix
approach\textquotedblright\ of random-matrix theory, this is provided by the
\textquotedblleft stub model\textquotedblright .\cite{WavesRM} In this
model, a fictitious \textquotedblleft stub\textquotedblright\ is attached to
the quantum dot, see Fig.\ \ref{fig:1a}. If the number of channels in the
stub is $M-N$, the scattering matrix $S$ can be written 
\begin{equation}
S=PU(1-Q^{\dagger }RQU)^{-1}P^{\dagger },  \label{eq:SU}
\end{equation}%
where $U$ is the $M\times M$ scattering matrix of the quantum dot (without
stub), $R$ is the $(M-N)$-dimensional reflection matrix of the stub, and $P$
and $Q$ are $N\times M$ and $(M-N)\times M$ projection matrices,
respectively, with $P_{ij}=\delta_{i,j}\openone$, and $Q_{ij}=\delta
_{i+N,j}\openone$. (Here $\openone$ is the $2\times 2$ unit matrix in spin
space.) The area and width of the stub are chosen such that (i) the dwell
time in the stub is much larger than the dwell time in the quantum dot and
(ii) the time for ergodic exploration of the dot plus stub system is much
shorter than the time for escape into one of the two leads. The first
condition ensures that all dependence on Fermi energy, magnetic field, or
spin-orbit scattering rate will be through the reflection matrix $R$, so
that the matrix $U$ can be taken at a fixed reference energy $\varepsilon =0$
and at zero magnetic field and spin-orbit scattering rate. Since scattering
from the quantum dot is chaotic, this implies that elements of $U$ are
proportional to the $2\times 2$ unit matrix in spin space $\openone$ and
that $U$ can be chosen from Dyson's circular orthogonal ensemble of random
matrix theory.\cite{Beenakker97} The second condition, which requires $M\gg
N $, ensures that electrons explore the phase space of the combined dot plus
stub system ergodically before they exit to the leads, so that the
distribution of the scattering matrix $S$ remains universal and is
unaffected by the addition of the stub. Indeed, the parametric dependence of
the scattering matrix described by the stub model has been shown to be the
same as that described by the Hamiltonian approach of random-matrix 
theory.\cite{WavesRM}

We take the reflection matrix $R$ of the form 
\begin{equation}
R=\mbox{exp}\left[ \frac{2\pi i}{M\Delta }(\varepsilon -
\mathcal{{H^{\prime }})}\right] ,  \label{eq:R}
\end{equation}%
where $\Delta $ is the mean level spacing of the closed dot and 
$\mathcal{{H^{\prime }}}$ is an $(M-N)$ dimensional matrix which 
describes the effect
of the magnetic field and spin-orbit scattering, 
\begin{eqnarray}
{\mathcal{H}^{\prime }} &=&\frac{\Delta }{2\pi }
  \left[i \mathcal{X}(x\openone+a_{\bot }\sigma_{3})+
  ia\left( \mathcal{A}_{1}\sigma_{1}+\mathcal{A}_{2}\sigma_{2}\right) 
  \right.  \notag \\
&&\left. \mbox{}-\mathbf{b}\cdot \mbox{\boldmath $\sigma$}+b_{\bot }
  \mathcal{B}_{h}\sigma_{3})\right] .
\end{eqnarray}%
Here $\mathcal{X}$, $\mathcal{A}_{1}$ and $\mathcal{A}_{2}$ are real
antisymmetric matrices, with $\left\langle \mbox{tr}\,
\mathcal{XX}^{T}\right\rangle =M^{2}$ and $\left\langle \mbox{tr}\,
\mathcal{A}_{i}
\mathcal{A}_{j}^{T}\right\rangle =\delta_{ij}M^{2}$, while $\mathcal{B}_{h}$
is a real symmetric matrix with $\left\langle \mbox{tr}\,\mathcal{B}
_{h}^{2}\right\rangle =M^{2}$. The dimensionless parameters $x$, $a_{\bot }$
, $a$, $b$, and $b_{\bot }$ correspond to the energy scales defined in
Eq.\ (\ref{eq:E_so}) as 
\begin{eqnarray}
x^{2} &=&\pi \varepsilon_{B}/\Delta ,  \notag \\
a_{\bot }^{2} &=&\pi \varepsilon_{\bot }^{\mbox{so}}/\Delta ,  \notag \\
a^{2} &=&\pi \varepsilon_{\Vert }^{\mbox{so}}/\Delta ,  \label{eq:defs} \\
b &=&\pi \varepsilon ^{\mathrm{Z}}/\Delta ,  \notag \\
b_{\bot }^{2} &=&\pi \varepsilon_{\bot }^{\mathrm{Z}}/\Delta .  \notag
\end{eqnarray}
We define the corresponding quantities $x^{\prime }$, $b^{\prime }$,
and $b_{\bot }^{\prime }$ when calculating a correlator between 
scattering matrix
elements at different values $\mathbf{B}$ and $\mathbf{B}^{\prime }$ of the
magnetic field. For weak uniform spin-orbit scattering, one has the
strong inequalities $a \ll a_{\bot}$, $b_{\bot} \ll b$.

To calculate the correlator (\ref{eq:corr}), we expand $S$ in powers of $U$
using Eq.~(\ref{eq:SU}) and integrate $U$ over the space of unitary
symmetric matrices using the diagrammatic technique of Ref.\
\onlinecite{diagrams}. To leading order in $1/M$ and $1/N$ we can take the
elements of $U$ to be Gaussian random variables with mean zero and with
variance $\left\langle U_{ij}U_{kl}^{\ast }\right\rangle =M^{-1}\left(
\delta_{ik}\delta_{jl}+\delta_{il}\delta_{jk}\right) $. Then only ladder
diagrams and maximally crossed diagrams contribute to the correlator
Eq.\ 
(\ref{eq:corr}). Summing the contributions from these two classes of
diagrams yields 
\begin{eqnarray}
\langle S_{kl;\mu \nu }^{\vphantom{*}}(\varepsilon ,\mathbf{B})S_{k^{\prime
}l^{\prime };\mu ^{\prime }\nu ^{\prime }}^{\ast }(\varepsilon ^{\prime },
\mathbf{B}^{\prime })\rangle \! &=&\delta_{kk^{\prime }}\delta_{ll^{\prime
}}D_{\mu \nu ;\nu ^{\prime }\mu ^{\prime }} \\
&&\!\mbox{}+\delta_{kl^{\prime }}\delta_{lk^{\prime }}(\mathcal{T}
C \mathcal{T})_{\mu \nu ;\mu ^{\prime }\nu ^{\prime }},  \notag
\end{eqnarray}%
where 
\begin{eqnarray}
D &=&\left( M\openone\otimes \openone-\mbox{tr}\,R\otimes R^{\prime
}{}^{\dagger }\right) ^{-1},  \notag \\
C &=&\left( M\openone\otimes \openone-\mbox{tr}\,R\otimes R^{\prime
}{}^{\ast }\right) ^{-1}.  \label{eq:CD_defs}
\end{eqnarray}%
Here $R^{\prime }$ is given by Eq.\ (\ref{eq:R}), with $x$, $b$, $b_{\bot }$
and $\varepsilon $ replaced by $x^{\prime }$, $b^{\prime }$, $b_{\bot
}^{\prime }$ and $\varepsilon ^{\prime }$, respectively. The superscript $
\ast $ denotes the quaternion complex conjugate of $R^{\prime }$, 
$\mathcal{T}=\openone\otimes \sigma_{2}$, and the trace is taken 
over the channel
indices only. The tensor multiplication in Eq.\ (\ref{eq:CD_defs}) has a
reverse-order multiplication for the Pauli matrices in second place, 
\begin{equation}
(\sigma_{i}\otimes \sigma_{j})(\sigma_{i^{\prime }}\otimes \sigma
_{j^{\prime }})=(\sigma_{i}\sigma_{i^{\prime }})\otimes (\sigma
_{j^{\prime }}\sigma_{j}).  \label{eq:mult}
\end{equation}%
The two contributions $C$ and $D$ are the equivalents of cooperon and
diffuson in the conventional diagrammatic perturbation 
theory.\cite{SimonsAltshuler} Taking the limit $M\rightarrow \infty $ 
and using Eq.\ (\ref{eq:R}) we find 
\begin{eqnarray}
D^{-1} &=&N\openone\otimes \openone+\frac{2\pi i(\varepsilon -\varepsilon
^{\prime })}{\Delta }\openone\otimes \openone  \notag  \label{eq:D} \\
&&\mbox{}+\frac{1}{2}[(x-x^{\prime })\openone\otimes \openone+a_{\bot
}(\sigma_{3}\otimes \openone-\openone\otimes \sigma_{3})]^{2}  \notag \\
&&\mbox{}+\frac{1}{2}a^{2}(\sigma_{1}\otimes \openone-\openone\otimes
\sigma_{1})^{2}  \notag \\
&&\mbox{}+\frac{1}{2}a^{2}(\sigma_{2}\otimes \openone-\openone\otimes
\sigma_{2})^{2}  \notag \\
&&\mbox{}+i\mathbf{b}\cdot (\mbox{\boldmath $\sigma$}\otimes \openone-
\openone\otimes \mbox{\boldmath $\sigma$})  \notag \\
&&\mbox{}+\frac{1}{2}(h\sigma_{3}\otimes \openone-h^{\prime }\openone
\otimes \sigma_{3})^{2}.
\end{eqnarray}%
\begin{eqnarray}
C^{-1} &=&N\openone\otimes \openone+\frac{2\pi i(\varepsilon -\varepsilon
^{\prime })}{\Delta }\openone\otimes \openone  \notag  \label{eq:C} \\
&&\mbox{}+\frac{1}{2}[(x+x^{\prime })\openone\otimes \openone+a_{\bot
}(\sigma_{3}\otimes \openone-\openone\otimes \sigma_{3})]^{2}  \notag \\
&&\mbox{}+\frac{1}{2}a^{2}(\sigma_{1}\otimes \openone-\openone\otimes
\sigma_{1})^{2}  \notag \\
&&\mbox{}+\frac{1}{2}a^{2}(\sigma_{2}\otimes \openone-\openone\otimes
\sigma_{2})^{2}  \notag \\
&&\mbox{}+i\mathbf{b}\cdot (\mbox{\boldmath $\sigma$}\otimes \openone+
\openone\otimes \mbox{\boldmath $\sigma$})  \notag \\
&&\mbox{}+\frac{1}{2}(h\sigma_{3}\otimes \openone+h^{\prime }\openone
\otimes \sigma_{3})^{2}.
\end{eqnarray}

\begin{figure}[t]
\epsfxsize=0.9\hsize 
\epsffile{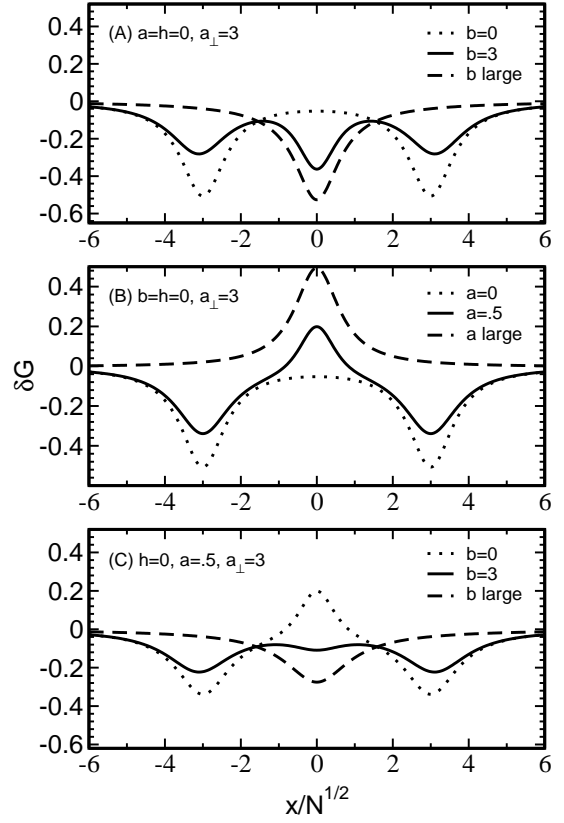} 
\caption{Dependence of the quantum-interference correction to the
conductance, $\protect\delta G$, on the perpendicular magnetic field $x$.
The conductance is measured in units of $(2e^2/h)N_{1}N_{2}/(N_1+N_2)^2$.
Panel (a) shows $\protect\delta G$ for $a_{\bot}=3$, $a=0$, $b_{\bot}=0$ and
two values of the parallel magnetic field, $b=0$ and $b=3$, and for the
limit $b \to \infty$. Panel (b) shows $\protect\delta G$ for
$a_{\bot}=3$, 
$b=b_{\bot}=0$, and three values of the spin-orbit parameter $a$:
$a=0$, 
$a=0.5$ and $a\to \infty$. Panel (c) shows $\protect\delta G$ in the
crossover regime, with $a_{\bot}=3.0$, $a=0.5$, and for the same values of
the parallel magnetic field as in (a). In all three plots, we have set 
$b_{\bot} = 0$. Choosing $b_{\bot} \ll b^2$ is appropriate for small quantum
dots with $L \ll \protect\lambda_{1,2}$. A small nonzero value of $b_{\bot}$
has only a slight effect on the shape of the curves, its general effect
being to further reduce weak antilocalization. }
\label{fig:1}
\end{figure}

Setting $x^{\prime}=x$, $\mathbf{b}=\mathbf{b}^{\prime}=b \hat {e_1}$
and 
$b_{\bot}^{\prime}=b_{\bot}$, we find the average conductance, 
\begin{eqnarray}
\left\langle G\right\rangle &=&\frac{2e^{2}}{h}\frac{N_{1}N_{2}}{N_1+N_2} +
\delta G, \\
\delta G &=& -\frac{e^2}{h} \frac{N_1 N_2}{N_1+N_2} \sum_{\mu ,\nu }\left( 
\mathcal{T}C\mathcal{T}\right)_{\mu \nu ;\mu \nu}  \notag \\
&=& -\frac{e^2}{h} \frac{N_1 N_2}{N_1+N_2} \left( \frac{1}{4 a^2 + F_C}
\right.  \notag \\
&& \left. \mbox{} + \frac{4b^{2}+G_C^2+2G_{C}F_{C}-16 a_{\bot}^2 x^2}{
4G_{C}b^{2}+G_{C}^{2}F_{C}-16a_{\bot}^{2}F_{C}x^{2}} \right),
\label{eq:weak}
\end{eqnarray}
where we abbreviated 
\begin{eqnarray}
F_{C} &=&N+2\left(x^{2}+b_{\bot}^{2}\right)  \label{eq:fc} \\
G_{C} &=&N+2\left(x^{2}+ a^{2}+a_{\bot}^{2}\right).  \notag
\end{eqnarray}
The expression for the average conductance simplifies considerably when
either the perpendicular magnetic field or the parallel magnetic field is
zero. In the first case, $b=b_{\bot}=0$, one finds 
\begin{eqnarray}
\delta G &=& \frac{e^2}{h} \frac{N_1 N_2}{N_1+N_2} \left[\frac{1}{N + 2x^2}
- \frac{1}{4a^2 + N + 2x^2}\right.  \notag \\
&& \left.\mbox{} - \sum_{\pm} \frac{1}{2a^2 + N + 2(a_{\bot} \pm x)^2}
\right].
\end{eqnarray}
This simple result can be understood recalling that the spin-orbit coupling
term $a_{\bot}$ acts as an effective perpendicular field with opposite sign
for up and down spins. When the applied perpendicular field $x$ exactly
cancels this effective field, i.e., when $x=\pm a_{\bot}$, time-reversal
symmetry is ``restored'' for one spin direction, leading to a weak
localization-like correction to the conductance centered at $x = \pm
a_{\bot} $. A non-zero value of the spin-orbit coupling term $a$ leads to a
weak antilocalization peak at $x=0$, while a finite parallel magnetic fields
causes a weak-localization peak at zero perpendicular field, as is
illustrated in Fig.~\ref{fig:1}b and a, respectively. Both a parallel
magnetic field and the spin-orbit coupling term $a$ suppress the features at 
$x = \pm a_{\bot}$. (The case $a\gg a_{\bot}$ is applicable when the
spin-orbit coupling is not uniform, as discussed in the introduction.)
Figure \ref{fig:1}c shows an example for the dependence of the
conductance $G $ on the perpendicular field in the crossover regime
when all parameters 
($a_{\bot}$, $a$, and $b$) are important.

In the special case when the perpendicular magnetic field, $x$, is zero, we
find 
\begin{eqnarray}
\delta G &=& -\frac{e^2}{h} \frac{N_1 N_2}{N_1+N_2} \left[ \frac{1}{N + 2
a^2 + 2a_{\bot}^2} \right.  \notag \\
&& \left. \mbox{} - \frac{2 a^2 + 2a_{\bot}^2 - b_{\bot}^2} {4 b^2 + (N + 2
b_{\bot}^2)(N + 2 a^2 + 2a_{\bot}^2)} \right.  \notag \\
&& \left. \mbox{} + \frac{1}{N + 4 a^2 + 2 b_{\bot}^2} \right].
\end{eqnarray}
In the case of uniform spin-orbit scattering, this may be further simplified
using $a^{2}\ll a_{\bot}^2$ and $b_{\bot}^2\ll b$,\cite{AF} 
\begin{eqnarray}
\delta G &=& -\frac{e^2}{h} \frac{N_1 N_2}{N_1+N_2} \left[ \frac{1}{2
a_{\bot}^2 + N}\right.  \label{eq:G_x0} \\
&&\left. \mbox{} - \frac{1}{4 a^2 + 2 b_{\bot}^2 + N} +\frac{a_{\bot}^2}{4
b^2 + N^2 + 2 N a_{\bot}^2}\right].  \notag
\end{eqnarray}

At finite temperatures, dephasing will lead to a further suppression of weak
localization. Dephasing can be included in the theory presented here by the
introduction of a fictitious voltage probe.\cite{BuettikerVoltage,BM,BB}
This amounts to the replacement $N\rightarrow N+\hbar /\tau_{\phi }\Delta $
in the final results (\ref{eq:weak})--(\ref{eq:G_x0}), where $\tau_{\phi }$
is the dephasing time and $\Delta $ is the mean level spacing in the closed
quantum dot.\cite{BCH}

\section{Conductance fluctuations}

\label{sec:3}

Conductance fluctuations are described by the covariance of the conductance
at two different values $\varepsilon $ and $\varepsilon ^{\prime }$ of the
Fermi energy and at two different magnetic fields $\mathbf{B}$ and 
$\mathbf{B}^{\prime }$, 
\begin{eqnarray}
  \mbox{cov}\,[G(\varepsilon ,\mathbf{B}),
  G(\varepsilon ^{\prime },\mathbf{B}^{\prime })] &=&
  \langle G(\varepsilon ,\mathbf{B}),G(\varepsilon ^{\prime },
  \mathbf{B}^{\prime })\rangle  \notag  \label{eq:cov_def} \\
  &&\mbox{}-\langle G(\varepsilon ,\mathbf{B})\rangle \langle 
  G(\varepsilon^{\prime },\mathbf{B}^{\prime })\rangle .  \notag \\
&&
\end{eqnarray}%
For this calculation, we need to know the average of a product of four
scattering matrix elements. However, if the conductance is expressed in
terms of the scattering matrix $S$ using Eq.\ (\ref{eq:Landauer}), to
leading order in $1/N$, the scattering matrix elements may be considered
Gaussian random numbers, and the average of four scattering matrix elements
can be factorized into products of pair averages of the form of Eq.\ 
(\ref{eq:corr}).\cite{Argaman} One thus obtains 
\begin{eqnarray}
\mbox{cov}\,\left[ G\left( \varepsilon ,\mathbf{B}\right) ,G\left(
\varepsilon ^{\prime },\mathbf{B}^{\prime }\right) \right] &=&\left( 
\frac{e^{2}}{h}\right) ^{2}\left( \frac{N_{1}N_{2}}{N_{1}+N_{2}}\right)^{2} 
\notag \\
&&\mbox{}\times \left( V_{D}+V_{C}\right) ,  \label{eq:covresult}
\end{eqnarray}
where 
\begin{widetext}
\begin{eqnarray} 
  \label{eq:D_def}
  V_D &=& \sum_{\mu,\nu,\mu',\nu' =\pm}
  D_{\mu \nu ;\mu' \nu'}D_{\nu' \mu';\nu \mu }
  \\
  &=& 
  \frac{\Xi_{D}}{\left| (b^{2}-b^{\prime
}{}^{2})^{2}+2K_{D}(-4a^{2}bb^{\prime }+(b^{2}+b^{\prime
}{}^{2})L_{D})+(L_{D}^{2}-4a^{4})(K_{D}^{2}-4a_{\bot}^{2}(x-x^{\prime
})^{2})\right| ^{2}}, \nonumber \\
  \label{eq:C_def}
  V_C &=& \sum_{\mu,\nu,\mu',\nu' =\pm}
  \left( {\cal T}C{\cal T}\right)_{\mu \nu ;\mu' \nu'}
  \left( {\cal T}C{\cal T}\right)_{\mu' \nu' ;\mu \nu}
  \\ &=& \nonumber
  \frac{\Xi_{C}}{\left| (b^{2}-b^{\prime
}{}^{2})^{2}+2K_{C}(4a^{2}bb^{\prime }+(b^{2}+b^{\prime
}{}^{2})L_{C})+(L_{C}^{2}-4a^{4})(K_{C}^{2}-4a_{\bot}^{2}(x+x^{\prime
})^{2})\right| ^{2}},
\end{eqnarray}
Here we abbreviated 
\begin{eqnarray*}
\Xi_{D} &=& 
  2\left| K_{D}(4a^{4}-L_{D}^{2})+
  4a^{2}bb^{\prime }-L_{D}(b^{2}+b^{\prime}{}^{2})\right| ^{2}
  + 2\left| K_{D}(b^{2}+b^{\prime}{}^{2})+K_{D}^{2}L_{D}-
  4a_{\bot}^{2}L_{D}(x-x^{\prime })^{2}\right|^{2}\\
  && \mbox{} + 
  8a_{\bot}^{2}(x-x^{\prime })^{2}\left(\left(b+b^{\prime }\right)^2+
  \left|L_{D}+2a^2\right|^2\right)\left(\left(b-b^{\prime }\right)^2+
  \left|L_{D}-2a^2\right|^2\right)
  +8\left| K_{D}bb^{\prime }+a^{2}(K_{D}^{2}-
  4a_{\bot}^{2}(x-x^{\prime})^{2})\right| ^{2}
  \\ && \mbox{} 
  + 4\left| b(b^{2}-b^{\prime }{}^{2})-K_{D}(2a^{2}b^{\prime}- 
  bL_{D})\right|^{2}+4\left| b^{\prime }(b^{\prime }{}^{2}-b^{2})-
  K_{D}(2a^{2}b-b^{\prime}L_{D})\right| ^{2} 
  + 8\left| a^{2}(b^{2}+b^{\prime }{}^{2})-
  bb^{\prime }L_{D}\right|^{2} \\
  \Xi_{C} &=& 2\left| K_{C}(4a^{4}-L_{C}^{2})-
  4a^{2}bb^{\prime }-L_{C}(b^{2}+b^{\prime}{}^{2})\right| ^{2}
  + 2\left| K_{C}(b^{2}+b^{\prime}{}^{2})+K_{C}^{2}L_{C}-
  4a_{\bot}^{2}L_{C}(x+x^{\prime })^{2}\right|^{2}\\
  && \mbox{} 
  + 8a_{\bot}^{2}(x+x^{\prime })^{2}\left(\left(b-b^{\prime }\right)^2+
  \left|L_{C}+2a^2\right|^2\right)\left(\left(b+b^{\prime }\right)^2+
  \left|L_{C}-2a^2\right|^2\right)
  + 8\left| K_{C}bb^{\prime }+a^{2}(-K_{C}^{2}+
  4a_{\bot}^{2}(x+x^{\prime})^{2})\right| ^{2}
  \\ && \mbox{} 
  + 4\left| b(b^{2}-b^{\prime }{}^{2})+
  K_{C}(2a^{2}b^{\prime} + bL_{C})\right|^{2}
  +4\left| b^{\prime }(b^{\prime }{}^{2}-b^{2})+
  K_{C}(2a^{2}b+b^{\prime}L_{C})\right| ^{2}
  + 8\left| a^{2}(b^{2}+b^{\prime }{}^{2})+
bb^{\prime }L_{C}\right|^{2}
\end{eqnarray*}
and 
\begin{eqnarray*}
  K_{D} &=& N + \frac{2\pi i ( \varepsilon -\varepsilon')}{\Delta} 
  + \frac{1}{2}\left(x-x^{\prime}\right)^2+
2\left( a^{2}+a_{\bot}^{2}\right)+\frac{1}{2}\left(b_{\bot}+
b_{\bot}^{\prime}\right)^2, \\
  L_{D} &=& N + \frac{2\pi i ( \varepsilon -\varepsilon')}{\Delta} 
  +\frac{1}{2}\left(x-x^{\prime}\right)^2+
2a^{2}+\frac{1}{2}\left(b_{\bot}-b_{\bot}^{\prime}\right)^2, \\
  K_{C} &=& N + \frac{2\pi i ( \varepsilon -\varepsilon')}{\Delta} 
  +\frac{1}{2}\left(x+x^{\prime}\right)^2+
  2\left( a^{2}+a_{\bot}^{2}\right)+\frac{1}{2}\left(b_{\bot}-
  b_{\bot}^{\prime}\right)^2, \\
  L_{C} &=& N + \frac{2\pi i ( \varepsilon -\varepsilon')}{\Delta} 
  +\frac{1}{2}\left(x+x^{\prime}\right)^2+
  2a^{2}+\frac{1}{2}\left(b_{\bot}+b_{\bot}^{\prime}\right)^2.
\end{eqnarray*}

\subsection{Variance}

The variance of the conductance at zero temperature is obtained from
Eq.\ (\ref{eq:covresult}) by setting $\varepsilon' = \varepsilon$ and
$\mathbf{B'} = \mathbf{B}$,
\begin{eqnarray}
  \mbox{var}\, G &=&
  \frac{e^{4}}{h^{2}} \frac{N_{1}^{2}N_{2}^{2}}{(N_1+N_2)^2}
  \left[\frac{1}{N^2} + \frac{1}{G_D^2} +
  \frac{8 b^2 + G_D^2 + (4 a^2 + N)^2}{(4 a^2 G_D + 4 b^2 + G_D N)^2}
  + \frac{1}{(4 a^2 + F_C)^2}
  \right. \nonumber \\ && \left. \mbox{}
  + \frac{2 F_C}{G_C (4 b^2 + G_C F_C) - 16 a_{\bot}^2 F_C x^2}
  + \frac{(4 b^2 + G_C^2 + 16 a_{\bot}^2 x^2)^2
    + 64 a_{\bot}^2 x^2 (F_C^2 - G_C^2)}
  {(G_C (4 b^2 + G_C F_C) - 16 a_{\bot}^2 F_C x^2)^2} \right].
  \label{eq:genvar}
\end{eqnarray}
where $F_{C}$ and $G_{C}$ are given by Eq.\ (\ref{eq:fc}), and
\begin{equation}  
  G_{D} =N+2\left( a^{2}+a_{\bot}^{2}+b_{\bot}^{2}\right).
\end{equation}
Simplifications occur in the limits of zero parallel or
perpendicular field, and for large perpendicular field. In the
absence of a parallel field, $b=b_{\bot}=0$, one finds the 
variance
\begin{eqnarray}
\mbox{var}\,  G &=& \frac{e^{4}}{h^{2}} \frac{N_{1}^{2}N_{2}^{2}}
  {(N_1 + N_2)^{2}}
\left[\frac{1}{(2 a^2 + N + 2 (a_{\bot} - x)^2)^2} + 
\frac{1}{(N + 2 x^2)^2} + \frac{1}{(4 a^2 + N + 2 x^2)^2}
\right.\nonumber \\ && \left. \mbox{}
  + \frac{1}{(2 a^2 + N + 2 (a_{\bot} + x)^2)^2}
  + \frac{1}{N^2} + \frac{1}{(4 a^2 + N)^2} + \frac{2}{(2 (a^2 +
a_{\bot}^2) + N)^2}\right],\label{eq:var_b0}
\end{eqnarray}
while in the absence of a perpendicular magnetic field, $x=0$, one has
\begin{eqnarray}
\mbox{var}\, G &=&\frac{e^{4}}{h^{2}} \frac{N_{1}^{2}N_{2}^{2}}
  {(N_1 + N_2)^{2}}\left[
  \frac{1}{N^2} + \frac{1}{(4 a^2 + 2 b_{\bot}^2 + N)^2} + \frac{2}{[2
  (a^2 + a_{\bot}^2 + b_{\bot}^2) + N]^2} 
  + \frac{2}{4 b^2 + (4 a^2 + N) [2 (a^2 + a_{\bot}^2) + N]}
  \nonumber \right. \\ && \left. \mbox{}
   + \frac{2}{4 b^2 + N (2 (a^2 + a_{\bot}^2) + N)} + 
  \frac{4 (a^2 + a_{\bot}^2)^2}
  {\left(4 b^2 + N [2 (a^2 + a_{\bot}^2) + N]\right)^2} 
  + \frac{4 (a^2 - a_{\bot}^2)^2}
  {\left(4 b^2
+ (4 a^2 + N) [2 (a^2 + a_{\bot}^2) + N]\right)^2}\right].
\label{eq:var_x0}
\end{eqnarray}
In the last equality we also used $b_{\bot} \ll b^2$. 
In the limit of large perpendicular field, $x \gg 1$, one has
\begin{eqnarray}
  \mbox{var}\, G &=&\frac{e^{4}}{h^{2}} \frac{N_{1}^{2}N_{2}^{2}}
  {(N_1 + N_2)^{2}}\left[\frac{1}{N^2}+ 
  \frac{2}{4b^2 + (4a^2 + N)\left[2(a^2 + a_{\bot}^2 + b_{\bot}^2) + N\right]}
  \nonumber \right.\\
  && \left. \mbox{} +
  \frac{1}{\left[2(a^2 + a_{\bot}^2 + b_{\bot}^2) + N\right]^{2}}
  + \frac{4(-a^2 + a_{\bot}^2 + b_{\bot}^2)^2}
  {\left(4b^2 + (4a^2 + N)\left[2(a^2 + a_{\bot}^2 + b_{\bot}^2) +
  N\right]\right)^2}
  \right].
\end{eqnarray}
When
both the perpendicular and the parallel field are large, i.e. 
$x^{2},b\gg a_{\bot}^{2},a^{2},b_{\bot}^{2},N$ a particularly simple 
expression for the variance results 
\begin{eqnarray}
\nonumber
\mbox{var}\, G &=& \frac{e^{4}}{h^{2}}
\frac{N_{1}^{2}N_{2}^{2}}{(N_1+N_2)^{2}}\left[
\frac{1}{N^2} 
  + \frac{1}{\left(N + 2a^2 + 2a_{\bot}^2 
  + 2b_{\bot}^2\right)^{2}}\right].\label{eq:var_large}
\end{eqnarray}
\end{widetext}

\begin{figure}[t]
\epsfxsize=0.9\hsize 
\epsffile{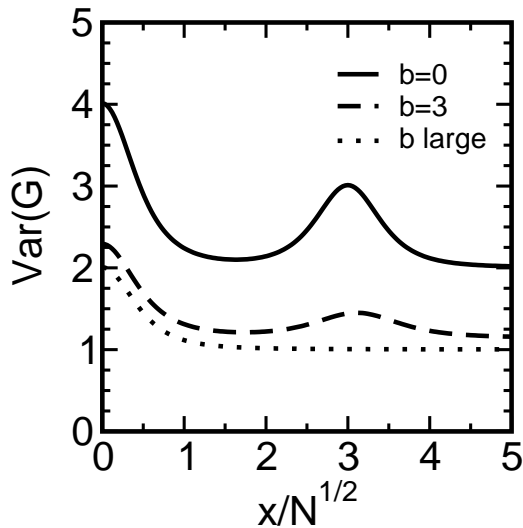} 
\caption{Conductance fluctuations $\mbox{var}\, G$, measured in units
  of 
$[(e^2/h)(N_1 N_2)/(N_1 + N_2)^2]^2$ as a function of the perpendicular
magnetic field $x$, for $a_{\bot}=3$, $a=0$, $b_{\bot}=0$. From bottom to
top, the curves are for $b=0$, $b=3$, and for the limit $b \to \infty$.}
\label{fig:2}
\end{figure}

The variance of the conductance in the presence of spin-orbit coupling is
shown in Fig.~\ref{fig:2}, for three different parallel field strengths. An
increase of the conductance fluctuations is observed around $x= \pm 
a_{\bot}$. This feature is explained noting that at $x=\pm a_{\bot}$ the
perpendicular magnetic field exactly cancels the effective magnetic field
due to the spin-orbit coupling term $a_{\bot}$ for one spin direction, so
that the effective Hamiltonian for that spin direction is real, not complex.
A parallel field and the spin-orbit coupling term $a$ suppress this feature
(data for $a$ not shown). Both effects also reduce the variance at zero
perpendicular field by a factor of two.

At finite temperatures, both the effects of thermal smearing and dephasing
need to be taken into account for a calculation of the variance of the
conductance. Dephasing (with dephasing time $\tau_{\phi }$) is incorporated
in the results (\ref{eq:covresult})--(\ref{eq:var_large}) by the
substitution $N\rightarrow N+\hbar /\tau_{\phi }\Delta $. The effect of
thermal smearing requires an integration over the energies $\varepsilon $
and $\varepsilon ^{\prime }$, 
\begin{equation}
\mbox{var}\,G\left( T\right) =
  \int d\varepsilon d\varepsilon ^{\prime }
  \frac{df}{d\varepsilon }\frac{df}{d\varepsilon ^{\prime }}
  \mbox{cov}\left[G(\varepsilon ),G(\varepsilon ^{\prime })\right] ,
  \label{eq:T}
\end{equation}%
where $f$ is the Fermi distribution function. The integrations over energy
needs to be done by numerical methods in all but a few special cases. We
refer to Ref.\ \onlinecite{thesis} for details. Equation (\ref{eq:T}) was
used for a quantitative comparison of theory and experiment by Zumb\"{u}hl 
\emph{et al}.\cite{Zumbuehl}

\subsection{Symmetry with respect to perpendicular magnetic field inversion}

\label{sec:4}

The presence of spin-orbit coupling has interesting consequences for the
symmetry of the conductance under reversal of the perpendicular field.
Although, in general, the conductance does not change when the total
magnetic field is reversed, $b\rightarrow -b$, $b_{\bot }\rightarrow
-b_{\bot }$, and $x\rightarrow -x$, the conductance need not be symmetric
under reversal of the perpendicular component $x$ only. In order to study
the symmetry of the conductance under reversal of the perpendicular
component $x$ of the magnetic field, we consider the correlator 
\begin{equation}
Q(x)=\frac{\left\langle G(x)G(-x)\right\rangle -\langle G(x)\rangle
  ^{2}}
{\mbox{var}\,G(x)}
\end{equation}%
for $x^{2}\gg N$. Perfect symmetry under reversal of the perpendicular
component of the magnetic field only implies $Q(x)=1$, while the absence
of any correlations between $G(x)$ and $G(-x)$ implies $Q(x)=0$. 

Note that the invariance of the two-terminal conductance conductance 
under a complete reversal of the magnetic field implies that
$$
  \left\langle G(x,b)G(-x,b)\right\rangle =
  \left\langle G(x,b)G(x,-b)\right\rangle,
$$
so that $Q$ can also be defined as a correlator for conductances
at opposite values of the parallel magnetic field, keeping the
perpendicular field constant.

In the absence of an in-plane magnetic field, reversal of the
perpendicular magnetic field component is the same as reversal of the total
magnetic field, so that $G$ is trivially symmetric in $x$. In the absence
of spin-orbit scattering, the sole effect of the parallel magnetic field is
to lift spin degeneracy, and symmetry of $G$ follows from the fact that the
conductance for each spin species is symmetric in $x$. In the presence of
both spin-orbit scattering and an in-plane field, there is no general
symmetry that requires that $G$ is symmetric under reversal 
$B_{3}\rightarrow -B_{3}$, however the measure of observable
asymmetry depends upon to what cross-over symmetry class the generated
random matrix theory ensemble belongs.

\begin{figure}[tbp]
\epsfxsize=0.9\hsize 
\epsffile{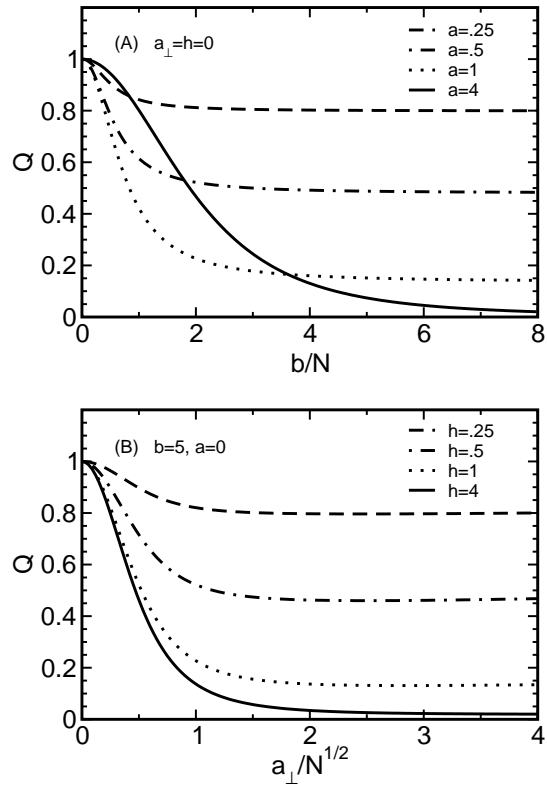} 
\caption{The correlator $Q(x)=(\left\langle G(x)G(-x)\right\rangle -\langle
G(x)\rangle ^{2})/\mbox{var}\,G(x)$ for $x^{2}\gg N$. In panel (a), $Q$ is
shown as a function of parallel field $b$, with $a_{\bot }=b_{\bot }=0$. In
panel (b), $Q$ is shown as a function of spin-orbit coupling $a_{\bot }$,
with $b=5$ and $a=0$.}
\label{fig:3}
\end{figure}

Indeed, calculation of the correlator $Q(x)$ using our general result 
(\ref{eq:covresult}) for the covariance of the conductance shows that the
symmetry of the conductance with respect to reversal of the perpendicular
component of the magnetic field is destroyed when either $b\neq 0$ and 
$a\neq 0$, $b_{\bot }\neq 0$ and $a_{\bot }\neq 0$, or $b_{\bot }\neq
0$ and 
$a\neq 0$. In the case $a=0$ and $b_{\bot }=0$, the conductance remains
symmetric in $x$, even if $a_{\bot }\neq 0$ and $b\neq 0$. In that case, the
effective Hamiltonian (\ref{eq:hamiltonian}) still has the extra
symmetry 
$\sigma_{1}\tilde{\mathcal{H}}\sigma_{1}=\tilde{\mathcal{H}}$, which
ensures the symmetry with respect to the perpendicular magnetic
field reversal.\cite{AF} Symmetry in the perpendicular component of the
magnetic field is also preserved if the parallel magnetic field is so large
that both $b$, $b_{\bot }\neq 0$, but $a_{\bot }^{2}\ll N$ and $a^{2}\ll N$,
see Ref.\ \onlinecite{AF} for a symmetry argument. Figure \ref{fig:3} shows
plots of $Q(x)$ for the two minimal cases $a=0$ or $b_{\bot }=0$. 

The general parameter dependence of $Q(x)$ for $x^{2}\gg N$ can be obtained
from Eq.\ (\ref{eq:covresult}). Simple expressions are found in the
cases 
$b\gg N$ and $a_{\bot }^{2}\gg N$, for which we have 
\begin{equation}
Q=\frac{N^{2}}{\left( 4a^{2}+2b_{\bot }^{2}+N\right) ^{2}},
\label{eq:theta_large}
\end{equation}%
and for $b\gg N$, while still $b_{\bot }=0$, 
\begin{eqnarray}
Q &=&1-\left[ 1-\frac{N^{2}}{(4a^{2}+N)^{2}}\right]  \\
&&\mbox{}\times \left[ 1-\frac{N^{2}}{2\left[ (a^{2}+a_{\bot
}^{2})^{2}+(a^{2}+a_{\bot }^{2}+N)^{2}\right] }\right] .  \notag
\end{eqnarray}%
On the other hand, if $b\gg N$ and $b_{\perp }\gg N$, the correlation
function $Q$ is given by 
\begin{equation}
Q=\frac{N^{2}}{\left( 2a^{2}+2a_{\bot }^{2}+N\right) ^{2}},
\label{eq:theta_bblarge}
\end{equation}%
confirming that $G$ remains a symmetric function of $x$ if $b$ and $b_{\perp
}$ are large but $a$ and $a_{\perp }$ are not.

Experimentally, the component of the conductance fluctuations 
that is antisymmetric with respect to the reversal
$B_{3}\rightarrow -B_{3}$ in the presence of in-plane magnetic 
field can be used as a test to the strength of spin-orbit coupling, 
even if $b\ll N$. This is
because an alternative mechanism of time-reversal symmetry breaking by an
in-plane magnetic field via the inter-subband mixing\cite{FJ} 
generates anti-symmetric conductance fluctuations in the third order
in the field only, thus 
$1-Q(x\gg N)\sim B_{\Vert }^{6}$. In contrast, the spin-orbit-coupling 
induced asymmetry, which can be analyzed by expanding $Q(x)$
with respect to $b^{2}$ and $b_{\bot }^{2}$, has
\begin{widetext}
\begin{eqnarray}
  Q &=& 1 - 4 [4 a^2 b^2 (32 a^4 a_{\perp}^4 + 24 a^2 a_{\perp}^4 N
  + 6 a_{\perp}^4 N^2 + 4 a_{\perp}^2 N^3 + N^4) 
  \nonumber \\ && \mbox{}
  + a_{\perp}^2 b_{\perp}^2 (128 a^6 a_{\perp}^4 + 96 a^4 a_{\perp}^4 N
  + 24 a^2 a_{\perp}^4 N^2 + 4 a_{\perp}^4 N^3 + 6 a_{\perp}^2 N^4 +
  3 N^5)]  \nonumber \\ && \mbox{} \times
  [N ( 4 a^2 + N ) (2 a_{\perp}^2 + N)
  (16 a^4 a_{\perp}^4 + 8 a^2 a_{\perp}^4 N + 2 a_{\perp}^4 N^2 + 
  2 a_{\perp}^2 N^3 + N^4)]^{-1},
\end{eqnarray}
\end{widetext}
for the case of uniform spin-orbit coupling. In the limit of a dot
with a large number of channels $N \gg a_{\perp}^2$ but still
in the universal regime $N \ll E_{\rm Th}/\Delta$, this simplifies
to
\begin{equation}
  Q = 1 - 16 a^2 b^2 N^{-3}.
\end{equation}
Replacing the dimensionless spin-orbit coupling and magnetic
field $a_{\perp}$ and $b_{\perp}$ by their dimensionful
counterparts, one arrives at the result (\ref{eq:AsymmetricPart}) 
advertised in the introduction.

\subsection{Autocorrelation function and the ``correlation echo''}

Correlations between the conductance $G$ at different values of the
perpendicular magnetic field $x$ are studied through the conductance
autocorrelation function 
\begin{equation}
\chi(x,x^{\prime}) = \left<G(x)G(x^{\prime})\right> - \langle G(x) \rangle
\langle G(x^{\prime}) \rangle.
\end{equation}
For large perpendicular fields, $x^2, x^{\prime}{}^2 \gg N$, the
autocorrelator $\chi(x,x^{\prime})$ depends on the difference 
$(x-x^{\prime}) $ only. Since time-reversal symmetry is fully broken for such
large magnetic fields, any features in the autocorrelator are signatures of
the spin-orbit coupling.

For zero parallel field, $b=b_{\bot}=0$, one finds from Eq.\ 
(\ref{eq:covresult}), 
\begin{eqnarray}
\chi(\xi) &=& \left(\frac{2 e^2}{h} \frac{N_1 N_2}{N_1 + N_2} \right)^2 
\left[\frac{1}{(2N + 8a^2 + \xi^2)^{2}}\right.  \notag \\
&& \mbox{} + \frac{1}{(2N + \xi^2)^{2}} + \frac{1}{(2N + 4a^2 + (-2a_{\bot}
+ \xi)^2)^{2}}  \notag \\
&& \left. \mbox{} + \frac{1}{(2N + 4a^2 + (2a_{\bot} + \xi)^2)^{2}}\right],
\label{eq:chi_b0}
\end{eqnarray}
where $\xi=x-x^{\prime}$.

\begin{figure}[tbp]
\epsfxsize=0.8\hsize 
\hspace{0.1\hsize} 
\epsffile{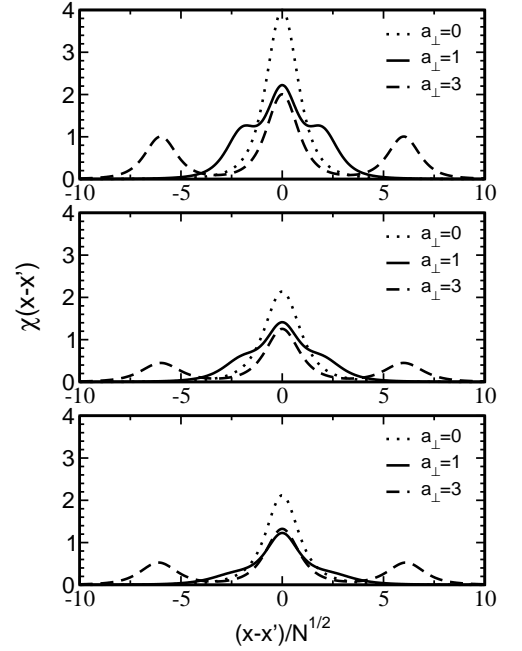} 
\caption{The correlator $\protect\chi(x-x^{\prime})$ as a function of the
dimensionless perpendicular magnetic field difference $(x-x^{\prime})$, for
three values of the spin-orbit term $a_{\bot}$: $a_{\bot}=0$, $a_{\bot}=1$,
and $a_{\bot}=3$. Curves in panel (a) are for $a=0$ and $b=b_{\bot}=0$.
Curves in panel (b) are for $a=0.5$ and $b=b_{\bot}=0$. Panel (c) shows the
effect of the parallel field, $b=2$, with $a=b_{\bot}=0$. In all three
panels, $\protect\chi(x-x^{\prime})$ is measured in units of $[(e^2/h)
N_{1}^{2}N_{2}^{2}/(N_1+N_2)^{2}]^2$. }
\label{fig:4}
\end{figure}

A plot of $\chi (x-x^{\prime })$ for different values of $a_{\bot }$ is
shown in Fig.~\ref{fig:4}. Remarkably, the conductance autocorrelator shows
an \textquotedblleft echo\textquotedblright\ at $|x-x^{\prime }|=
2a_{\bot }$. This echo appears even for large values of $a_{\bot }$, 
for which the
conductance correlations have decayed for intermediate values of the
magnetic-field difference, see Fig.\ \ref{fig:4}a. The magnetic-field echo
has the same origin as the reappearance of the weak localization and
conductance fluctuations at finite values of the perpendicular magnetic
field $x$: The spin-orbit term $a_{\bot }$ acts as a homogeneous
perpendicular field with a different direction for the up and down spins
(defined with respect to the axis $\mathbf{\hat{e}}_{3}$ perpendicular to
the plane of the quantum well). In the presence of an external perpendicular
magnetic field $x$, Up spins and down spins experience effective
perpendicular magnetic fields $x+a_{\bot }$ and $x-a_{\bot }$, respectively.
If the magnetic field difference $x-x^{\prime }$ equals $\pm 2a_{\bot }$,
one spin species at field $x$ moves in the same effective perpendicular
field as the other spin species at field $x^{\prime }$. Hence, in the
absence of a parallel field and a spin-orbit term $a$, the conductance has
identical fluctuating contributions at $x$ and $x^{\prime }$. However, only
one spin species is involved in the \textquotedblleft
echo\textquotedblright\ --- the autocorrelation function reaches only one
half of the total variance of the conductance; the other spin species have
uncorrelated contributions to the conductance at the external fields $x$ and 
$x^{\prime }$. A magnetic field in the plane of the quantum well and the
spin-orbit coupling term $a$ suppress the echo, since they cause scattering
between the $\pm \mathbf{\hat{e}}_{3}$ spin directions, see Figs.\ 
\ref{fig:4}b and c.

\section{Effect of a non-uniform SO coupling.}
\label{sec:nonuniform}

The expressions of the previous two sections are valid for arbitrary
magnitudes of the dimensionless spin-orbit rates $a_{\perp}$ and 
$a$. In the case of uniform spin-orbit coupling, one has the
strong inequality $a \ll a_{\perp}$. In general, the spin-orbit 
coupling may be non-uniform throughout the quantum dot, {\em e.g.}, 
as a result of a spatial modulation of the asymmetry of the quantum 
well or as a result of a modulation of the electron density 
(and therefore confining potential width) across the dot. 
The main effect of a non-uniformity of the spin-orbit coupling 
is an increase of $a$,\cite{BCH} thus promoting the crossover into 
the standard symplectic ensemble. In this section, we'll give a
quantitative estimate of this effect for a special choice of the
non-uniformity. Another effect of a non-uniform spin-orbit coupling 
is a rescaling of $a_{\perp }$, and will not be considered here.

Assuming that the non-uniformity of spin-orbit coupling affects only
its strength without changing the generic form of the Hamiltonian, we
describe its effect via coordinate dependent spin-orbit coupling
lengths $\lambda_{1,2}^{-1}(\mathbf{r})$,
\begin{eqnarray}
\mathcal{H}\! &=&\!\frac{1}{2m}\left( \mathbf{p}-e\mathbf{A}-
  \mathbf{a}
  \right) ^{2}, \nonumber \\
  \mathbf{a} &=& \frac{\sigma_{2}}{2\lambda_{1}(\mathbf{r})}
  \mathbf{\hat{e_1}} -
  \frac{\sigma_{1}}{2\lambda_{2}(\mathbf{r})}
  \mathbf{\hat{e_2}}.
\end{eqnarray}%
In the case that 
$$
  \beta =\nabla \times \mathbf{a} =
  \frac{\sigma_{2}}{2}\partial_{x_{2}}\lambda_{1}^{-1}(\mathbf{r})
  + \frac{\sigma_{1}}{2}\partial_{x_{1}}\lambda_{2}^{-1}(\mathbf{r})
  \neq 0,
$$
it is not possible to eliminate the spin-orbit coupling from the
Hamiltonian in the first order in spin-orbit coupling constant by a unitary
transformation. Qualitatively, this explains why the spatial variation
of the spin-orbit coupling length gives rise of a term with the
symmetry of $\mathbf{a_{\parallel}}$. For a quantitative estimate of
how much the spatial variations of $\lambda_{1,2}$ contribute to the
energy scale $\varepsilon_{\parallel}$,
we note that it is still possible to find a transformation
$U$ such that the resulting non-Abelian vector potential in a locally
rotated spin frame,
\begin{equation}
\mathbf{\tilde{a}=}-iU^{\dagger }\nabla U-U^{\dagger }\mathbf{a}U,
\end{equation}%
would satisfy the conditions
\begin{equation}
  \nabla \cdot \mathbf{\tilde{a}} = 0
\end{equation}
in the interior of the quantum dot and $ \mathbf{\hat{n}}\cdot
\mathbf{\tilde{a}}=0$, on the dot boundary, where $\mathbf{\hat{n}}$
is the unit vector normal to the dot's boundary, at least, in the
lowest order in SO coupling. Such conditions would enable one to
satisfy boundary conditions for Cooperon and diffuson propagators from
diagrammatic perturbation theory, which is needed in order to make the
zero-dimensional approximation necessary for the use of random matrix
theory. This program is carried out in the appendix for a circular
quantum dot with a special choice of the coordinate dependence of the
spin-orbit coupling lengths $\lambda_{1,2}$.

Here, as an example, we consider here the case where the spin-orbit
coupling is stronger in the center of a quantum dot (where 
density is higher and the well is narrower) than
at the edges (where density is lower and the well is broader),
\begin{equation}
\frac{1}{\lambda_{1,2}}=\dfrac{1}{\lambda ^{\mathrm{c}}}-\frac{\left(
r/L\right) ^{2}}{\lambda ^{\mathrm{f}}},  \label{eq:nonunif}
\end{equation}%
where $L$ is the typical size of the dot, without making a 
restriction of the size of the dot. We perform a unitary
transformation
\begin{equation}
U = \exp \left( i\frac{x_{1}\sigma_{2}-x_{2}\sigma_{1}}
  {2\lambda^{\mathrm{c}}}\left[ 1-
  \dfrac{\lambda^{\mathrm{c}}}{3\lambda ^{\mathrm{f}}}
\left( \dfrac{r}{L}\right) ^{2}\right] \right),
\end{equation}
which rotates to a local spin reference axis.
With this transformation, the matrix vector $\mathbf{\tilde{a}}$
becomes, to lowest order in the inverse spin-orbit coupling length,
\begin{eqnarray}
\mathbf{\tilde{a}} &=& -i U^{\dagger} \nabla U - 
  \mathbf{a} + {\cal O}(\lambda^{-2}) \nonumber \\
&=&-\dfrac{x_{1}\sigma_{1}+x_{2}\sigma_{2}}{3\lambda ^{\mathrm{f}}L^2}\left[ 
\mathbf{\hat{e}}_{3}\times \mathbf{r}\right].
\end{eqnarray}%
We conclude that (a) the spin relaxation and the crossover into 
symplectic ensemble start, indeed, in the lowest order in the
non-uniform part of the spin-orbit coupling and (b) for the
non-uniform coupling (\ref{eq:nonunif}), the functional form of
the perturbation after transformation
is equal to that of the term $\mathbf{a_{\parallel}}$
in Eq.\ (\ref{eq:hamiltonian}). The latter observation allows us to
express the contribution of the non-uniformity to the energy scale
$\varepsilon_{\parallel}$ as
\begin{equation}
  \varepsilon_{\parallel} =
  \frac{1}{4} \kappa' \kappa E_{\rm Th}
  \left( \frac{L^2}{4 \lambda^{{\rm f}2}} \right),
  \label{eq:kappageneral}
\end{equation}
where the geometry-dependent coefficients $\kappa$ and $\kappa'$ are
the same as those defined for the spatially homogeneous spin-orbit
coupling in Eq.\ (\ref{eq:kappa}).

\section{Conclusions}

\label{sec:5}

In this paper we calculated the average and the fluctuations of the
conductance of a chaotic quantum dot in the presence of Bychkov-Rashba
and (linear) Dresselhaus spin-orbit coupling and a magnetic field and
studied the symmetry of the conductance under reversal of the 
perpendicular component of the magnetic
field. The calculations were done to leading order in $1/N$, where
$N=N_1+N_2$ is the total number of open channels in the leads
connecting the quantum dot to the reservoirs.  After a unitary
transformation, the spin-orbit term in the Hamiltonian and the Zeeman
coupling to the external magnetic field are transformed into four
terms that all have different symmetries.\cite{AF}
The corresponding energy scales
are the conventional Zeeman energy $\varepsilon^{\rm Z} = \mu_B g B$ and
three energy scales that depend on the spin-orbit coupling: two
spin-orbit scales $\varepsilon^{\rm so}_{\perp}$ and $\varepsilon^{\rm
so}_{\parallel}$, and the second Zeeman energy scale $\varepsilon^{\rm
Z}_{\perp}$. With the help of the theory presented here, the energy
scales $\varepsilon^{\rm so}_{\perp}$, $\varepsilon^{\rm
so}_{\parallel}$, and $\varepsilon^{\rm Z}_{\perp}$ can be obtained
from a measurement of the conductance autocorrelation function of the
quantum dot.

Our results provide several routes to a direct measurement of the
product of the spin-orbit coupling lengths $\lambda_1$ and 
$\lambda_2$. The product $\lambda_1 \lambda_2$ can be found from
a ``magnetic-field echo'' in the conductance fluctuations as a 
function of the perpendicular magnetic field. The magnetic-field
echo appears at the magnetic-field difference
\begin{equation}
  \Delta B = \frac{\hbar}{e \lambda_1 \lambda_1}.
\end{equation}
The same magnetic-field scale splits the weak-localization peak in the
absence of a parallel magnetic field. The ratio $\lambda_1/\lambda_2$
can be found from a measurement of the energy scale $\varepsilon^{\rm
Z}_{\perp}$ in a roughly circular quantum dot. In this case,
$(\lambda_2/\lambda_1)^2$ is equal to the ratio of the energy scales
$\varepsilon^{\rm Z}_{\perp}$ for the in-plane magnetic field in the
crystallographic directions $\hat e_1 = [110]$ and $\hat e_2 = [1\bar
1 0]$.  Measurement of $\varepsilon^{\rm so}_{\parallel}$ is less
suitable to determine $\lambda_1/\lambda_2$, as it may be
affected by non-uniformities in the spin-orbit scattering rate.

The theory presented here has been used for the interpretation of
measurements of the conductance of quantum dots in GaAs/GaAlAs 
heterostructures with spin-orbit coupling as a function of 
parallel and perpendicular components of the magnetic field
by Zumb\"{u}hl {\em et al.}\cite{Zumbuehl} 
The quantum dots used in the experiments of Ref.\
\onlinecite{Zumbuehl}
were built of the same material, but they varied in size, ranging 
from $L/\lambda=0.27$ to $0.64$.
Here $\lambda=(\lambda_{1}\lambda_{2})^{1/2}$ is the
geometric average of the spin-orbit lengths $\lambda_1$ and
$\lambda_2$. Whereas the conductance of smaller dots had a
minimum at zero perpendicular magnetic field, the conductance
of larger dots had a (local) maximum at zero magnetic field. 
{}From a fit to our expression for the average
conductance at zero parallel field, Eq. (\ref{eq:weak}), values for the
experimentally unknown parameters $\lambda =
4.4\mu \mathrm{m}$, the decoherence
time $\tau_{\phi }$ and a geometrical factor, see Eqs. (\ref{eq:E_so}), were
extracted. 
Measurement of the average conductance in
the presence of a magnetic field parallel to the plane of the
two-dimensional electron gas was in good agreement with our Eq.\
(\ref{eq:weak}), without additional fit parameters and
up to a parallel field of about 0.3 T.\cite{Zumbuehl} 
Upon inclusion of the effects of time-reversal symmetry breaking by a strong parallel field
due to the finite extent of the electron wavefunction across the
heterostructure,\cite{FJ} the authors of Ref.\ \onlinecite{Zumbuehl}
could extend the agreement between theory and experiment is 
extended to even higher magnetic fields.

\begin{acknowledgments}
We thank Igor Aleiner, Boris Altshuler,
Bertrand Halperin, Charles Marcus, Jeffrey Miller,
Yuli Nazarov, and Dominik Zumb\"uhl
for important discussions.
This work was supported by NSF under grant no.\ PHY 0117795
(JC), by the NSF under grant no.\ DMR
0086509 and by the Packard foundation (PWB), and
by EPSRC and NATO CLG (VF).
\end{acknowledgments}

\appendix

\section{Geometry-dependent coefficients for a circular quantum dot}

\label{app:1}

In this appendix, we calculate 
\begin{eqnarray}
\mathcal{D}_{\mu_{1}\nu_{1};\mu_{2}\nu_{2}}
  (\mathbf{r}_{1},\mathbf{r}_{2}) &=&
  \langle G_{\mu_{1}\nu_{1}}^{\mathrm{R}}
  (\mathbf{r}_{1},\mathbf{r}_{2})(E,B)  \notag \\
  &&\mbox{}\times G_{\mu_{2}\nu_{2}}^{\mathrm{A}}
  (\mathbf{r}_{1},\mathbf{r}_{2})(E^{\prime },B^{\prime })\rangle
\end{eqnarray}%
for a disordered circular quantum dot with spin-orbit scattering. Here 
$G^{\mathrm{R}}$ and $G^{\mathrm{A}}$ are the retarded and advanced Green
functions (inverses of $\varepsilon -\mathcal{H}$ for $\varepsilon $ just
above and just below the real axis, respectively), respectively, and the
indices $\mu_{1}$, $\mu_{2}$, $\nu_{1}$, and $\nu_{2}$ refer to the
electron spin. Comparison of our result with the random matrix theory of
Sec.\ \ref{sec:2} allows us to find the geometry-dependent
coefficients 
$\kappa $, $\kappa ^{\prime }$, and $\kappa ^{\prime \prime }$ of
Eqs.\ (\ref{eq:E_so}).

In a closed dot of volume $V$ and with diffusive dynamics, the
diffuson 
$\mathcal{D}(\mathbf{r}{_{1},\mathbf{r}_{2}})$ obeys the equation 
\begin{equation}
(i\hbar \omega +\Pi )\mathcal{D}(\mathbf{r}{_{1},\mathbf{r}_{2}})=\frac{2\pi 
}{V\Delta }\delta (\mathbf{r}_{1}-\mathbf{r}_{2})\openone\otimes
  \openone,
  \label{eq:diff}
\end{equation}%
where $\hbar \omega =\varepsilon ^{\prime }-\varepsilon $ and the
differential operator $\Pi $ is given by 
\begin{eqnarray}
\Pi &=&(D/\hbar )(-i\hbar \nabla_{\mathbf{r}_{1}}+\mathbf{a}\otimes
\openone
-\openone\otimes \mathbf{a}^{\prime })^{2}  \notag \\
&&\mbox{}+ih_{\mathrm{Z}}\otimes \openone-i\openone\otimes 
h_{\mathrm{Z}}^{\prime }.  \label{eq:Pidef}
\end{eqnarray}%
Here $D=v_{F}l/2$ is the diffusion coefficient in the quantum dot, the
matrix vector potentials $\mathbf{a}$ and $\mathbf{a}^{\prime }$ are the sum of a
spin-independent contribution from the perpendicular magnetic field and a
spin-dependent contribution from the Bychkov-Rashba and Dresselhaus 
spin-orbit
coupling, 
\begin{eqnarray}
\mathbf{a} &=&\left( \frac{eB_{3}y}{2c}-\frac{\hbar }{2\lambda_{1}}\sigma
_{2}\right) \mathbf{\hat{e}}_{1}-\left(
\frac{eB_{3}x}{2c}-\frac{\hbar}
{2\lambda_{2}}\sigma_{1}\right) \mathbf{\hat{e}}_{2},  \notag \\
\mathbf{a}^{\prime } &=&\left( \frac{eB_{3}^{\prime }y}{2c}-
\frac{\hbar }{2\lambda_{1}}\sigma_{2}\right) \mathbf{\hat{e}}_{1}-
\left( \frac{eB_{3}^{\prime }x}{2c}-\frac{\hbar }{2\lambda_{2}}
\sigma_{1}\right) 
\mathbf{\hat{e}}_{2},
  \nonumber \\
\end{eqnarray}%
while $h_{\mathrm{Z}}$ and $h_{\mathrm{Z}}^{\prime }$ represent the Zeeman
coupling to the magnetic field, 
$$
h_{\mathrm{Z}}=\frac{1}{2}\mu_{B}g\mathbf{B}\cdot \mbox{\boldmath $\sigma$},
\ \ 
h_{\mathrm{Z}}^{\prime }=\frac{1}{2}\mu_{B}g\mathbf{B}^{\prime }\cdot 
\mbox{\boldmath $\sigma$}.
$$
Equation (\ref{eq:diff}) is supplemented with the boundary condition 
\begin{equation}
\hat{n}\cdot (-i\hbar \nabla_{\mathbf{r}_{1}}+\mathbf{a}\otimes \openone-
\openone\otimes \mathbf{a}^{\prime })\mathcal{D}(\mathbf{r}{_{1},
\mathbf{r}_{2}})=0  \label{eq:boundary}
\end{equation}%
at the sample boundary, where $\hat{n}$ is the unit vector normal to the
boundary.

To find $\mathcal{D}$, we shift to the basis of eigenfunctions of the
diffusion equation (i.e., eigenfunctions of $\Pi $ in the absence of
spin-orbit scattering and a magnetic field) and calculate the matrix
elements of $\Pi $ in that basis. For this procedure to work, it is
important that the matrix vector potentials $\mathbf{a}$ and $\mathbf{a}^{\prime }$
do not change the Von Neumann boundary conditions (\ref{eq:boundary}), i.e.,
that 
\begin{equation}
\mathbf{\hat{n}}\cdot \mathbf{a}=\mathbf{\hat{n}}\cdot \mathbf{a}^{\prime }=0
\label{eq:r1}
\end{equation}%
at the boundary of the quantum dot. This requirement ensures that the
eigenfunctions of the diffusion equation with Von Neumann boundary
conditions can be used to construct the diffuson $\mathcal{D}$. In order to
fix the gauge, we also require that 
\begin{equation}
\nabla \cdot \mathbf{a}=\nabla \cdot \mathbf{a}^{\prime }=0  \label{eq:r2}
\end{equation}%
in the interior of the dot (London gauge).

As discussed in Ref.\ \onlinecite{AF} and in the introduction of this paper,
the requirements (\ref{eq:r1}) and (\ref{eq:r2}) are realized by performing
a suitable unitary transformation $U$ to the Hamiltonian, $\mathcal{H} \to
U^{\dagger} \mathcal{H} U$. For Eq.\ (\ref{eq:diff}), such a unitary
transformation implies 
\begin{eqnarray}
\mathcal{D} &\to& (U^{\dagger} \otimes U) \mathcal{D} (U \otimes
U^{\dagger}), \\
\Pi &\to& (U^{\dagger} \otimes U) \Pi (U \otimes U^{\dagger}).
\end{eqnarray}

We'll now perform an explicit calculation for the case of a circular quantum
dot of radius $L$, the origin of our coordinate system being chosen at the
center of the dot. In polar coordinates $x_{1}=r\cos \phi $, $x_{2}=r\sin
\phi $, the tensor eigenfunctions of the diffusion equation are of the form 
\begin{eqnarray}
f_{n0}^{ij}(r,\phi ) &=&c_{n0}J_{0}(rx_{n0}/L)\,\sigma_{i}\otimes \sigma
_{j},  \notag \\
f_{nm}^{ij}(r,\phi ) &=&c_{nm}J_{m}(rx_{nm}/L)\cos (m\phi )\,\sigma
_{i}\otimes \sigma_{j},  \notag \\
g_{nm}^{ij}(r,\phi ) &=&c_{nm}J_{m}(rx_{nm}/L)\sin (m\phi )\,\sigma
_{i}\otimes \sigma_{j},
\end{eqnarray}%
where $i,j=0,1,2,3$, $m=1,2,\ldots $, $x_{nm}$, $n=1,2,\ldots $, is
the 
$n$th positive root of $J_{m}^{\prime }(x)=0$, and we used the 
notation $\sigma
_{0}=\openone$. The corresponding eigenvalues are 
\begin{equation}
\lambda_{nm}=D\hbar (x_{nm}/L)^{2}
\end{equation}%
and the normalization constants are 
\begin{equation}
c_{nm}^{-2}=\frac{1}{2}\pi
L^{2}(J_{m}(x_{nm})^{2}-J_{m+1}(x_{nm})^{2})(1+\delta_{m0}).
\end{equation}%
The smallest eigenvalue is found for $n=m=0$, $\lambda_{00}=0$; all other
eigenvalues are larger than the Thouless energy $E_{\mathrm{Th}}=D\hbar
/L^{2}$.

For a circular quantum dot, the magnetic-field contribution to the matrix
vector potential of Eq.\ (\ref{eq:Pidef}) already satisfies the requirements
(\ref{eq:r1}) and (\ref{eq:r2}). For the spin-orbit contribution we use the
transformation 
\begin{eqnarray}
U &=& \exp \left[i \frac{x_1 \sigma_2}{2 \lambda_1} - i \frac{x_2
    \sigma_1}{2 \lambda_2} \right. \\
&& \left. \mbox{} - i \frac{(x_1^2 + x_2^2 - 3 L^2)}{48 \lambda_1 \lambda_2}
\left( \frac{x_2 \sigma_1}{\lambda_1} - \frac{x_1 \sigma_2}{\lambda_2}
\right) \right] ,  \notag
\end{eqnarray}
which fulfills the requirements (\ref{eq:r1}) and (\ref{eq:r2}) to leading
order in $L/\lambda$. [Note: although the transformation of 
Eq.\ (\ref{eq:U}) is sufficient to satisfy the requirements to order 
$L/\lambda$, the
remaining term $a_{\Vert}$ of Eq.\ (\ref{eq:hamiltonian}) does not obey Eq.\
(\ref{eq:r2}).]

With this transformation, the matrix vector potential $\mathbf{a}$ and the
Zeeman term $h_{\rm Z}$ change to 
\begin{eqnarray}
\mathbf{a} &\rightarrow &\mathbf{\hat{e}}_{1}\left[ \frac{eB_{3}x_{2}}{2c}+
\frac{\hbar x_{2}\sigma_{3}}{4\lambda_{1}\lambda_{2}}\right.  \notag \\
&&\left. \mbox{}+\frac{\hbar }{16\lambda_{1}\lambda_{2}}\left( 
\frac{2x_{1}x_{2}\sigma_{1}}{\lambda
 _{1}}+\frac{(3x_{2}^{2}+x_{1}^{2}-L^{2})
\sigma_{2}}{\lambda_{2}}\right) \right]  \notag \\
&&\mbox{}-\hat{e}_{2}\left[ \frac{eB_{3}x_{1}}{2c}+\frac{\hbar x_{1}\sigma
_{3}}{4\lambda_{1}\lambda_{2}}\right.  \notag \\
&&\left. \mbox{}+\frac{\hbar }{16\lambda_{1}\lambda_{2}}\left( 
\frac{2x_{1}x_{2}\sigma_{2}}{\lambda_{2}}+\frac{(3x_{1}^{2}+x_{2}^{2}-L^{2})
\sigma_{1}}{\lambda_{1}}\right) \right] ,  \notag \\
h_{\mathrm{Z}} &\rightarrow &\frac{1}{2}\mu_{B}g\mathbf{B}\cdot 
\mbox{\boldmath $\sigma$}-\frac{1}{2}\sigma_{3}\mu_{B}g\left(
\frac{B_{1}x_{1}}
{\lambda_{1}
}+\frac{B_{2}x_{2}}{\lambda_{2}}\right) ,  \label{eq:anew}
\end{eqnarray}%
with similar changes for $\mathbf{a}^{\prime }$ and $h^{\prime}_{\rm Z}$.

As long as all relevant energy scales are much smaller than 
$E_{\mathrm{Th}}$, it is sufficient to know the action of $\Pi $ on
the 16 eigenfunctions at $f_{00}^{ij}$ at $n=m=0$. In tensor notation, 
this action is 
\begin{widetext}
\begin{eqnarray}
  \Pi &=&
  \left. \frac{D \hbar}{2 L^2}
  \left(\frac{e}{c h} (\Phi - \Phi') \openone \otimes \openone
    + \frac{L^2}{4 \lambda^2} (\sigma_3 \otimes \openone -
      \openone \otimes \sigma_3) \right)^2 
  \right. \ \ \nonumber \\ && \left. \mbox{} +
  \frac{D \hbar L^6}{96 \lambda^4}
  \left( \frac{\sigma_1 \otimes \openone -
  \openone \otimes \sigma_1}{2 \lambda_1} \right)^2
  + \frac{D \hbar L^6}{96 \lambda^4}
  \left( \frac{\sigma_2 \otimes \openone -
  \openone \otimes \sigma_2}{2 \lambda_2} \right)^2
  \right. \ \ \nonumber \\ && \left. \mbox{} +
  i \frac{\mu_B g \mathbf{B}}{2} (\mbox{\boldmath $\sigma$} \otimes \openone
  - \openone \otimes \mbox{\boldmath $\sigma$}) 
  + \frac{\mu_B^2 g^2 L^4}{D \hbar } \kappa_h
    \left(\frac{B_1^2}{\lambda_1^2} + \frac{B_2^2}{\lambda_2^2}
    \right)
    (\openone \otimes \sigma_3 - \sigma_3 \otimes \openone)^2
    \right..
\end{eqnarray}
\end{widetext}where $\Phi =B_{3}\pi L^{2}$ is the magnetic flux through the
quantum dot and 
$$
\kappa_{h}=\frac{1}{2}\sum_{n=1}^{\infty}
\frac{1}{x_{1n}^{4}(x_{1n}^{2}-1)}
\approx 0.0182
$$
is a numerical constant. 
Confining ourselves to the case $\lambda =\lambda_{1}=\lambda_{2}$ and
comparing this result to Eq.\ (\ref{eq:D}), we can make the identifications 
\begin{eqnarray}
  x^{2} &=&\left( \frac{e\Phi }{ch}\right) ^{2}\frac{\hbar v_{F}l\pi }
  {L^{2}\Delta }, \\
  a_{\bot }^{2} &=&
  \left( \frac{L^{2}}{4\lambda^{2}}\right)^{2}\frac{\hbar
  v_{F}l\pi }{L^{2}\Delta }, \\
  a^{2} &=&\left( \frac{L^{2}}{4\lambda ^{2}}\right)^{3} \frac{v_{F}l\pi}
  {3L^{2}\Delta }, \\
  b &=&\frac{\mu_{B}g\pi B}{\Delta }, \\
  b_{\bot } &\approx &
  1.83\frac{L^{2}}{\hbar v_{F}l\Delta }(\mu_{B}Bg)^{2}
  \left( \frac{L^{2}}{4\lambda ^{2}}\right) .
\end{eqnarray}%
For the geometry-dependent constants $\kappa $, $\kappa ^{\prime }$, 
and $\kappa ^{\prime \prime }$ in Eq.\ (\ref{eq:E_so}), this implies 
$\kappa =2$, $\kappa ^{\prime }=1/3$ and $\kappa ^{\prime \prime }
\approx 0.292$.

If the spin-orbit coupling is not uniform throughout the quantum
dot, see Sec.\ \ref{sec:nonuniform}, the above estimates 
for $a_{\perp}$ and $a$ and the corresponding
energy scales $\varepsilon_{\bot}^{\rm so}$ and
$\varepsilon_{\Vert}^{\rm so}$ need to be revisited. Here we 
consider the example of Eq.\ (\ref{eq:nonunif}),
$\lambda^{-1}_{1,2}(\mathbf{r}) =
\lambda_{\rm c}^{-1} + (r^2/L^2 \lambda^{\rm f})$, and
construct the unitary transformation $U$ that ensures that the
conditions (\ref{eq:r1}) and (\ref{eq:r2}) are satisfied to 
lowest order in $1/\lambda_{f}$,
\begin{eqnarray}
  U &=& \exp \left[i \frac{x_1 \sigma_2}{2 \lambda_1^{\rm c}}
  - i \frac{x_2 \sigma_1}{2 \lambda_2^{\rm c}}
  \right. \nonumber \\ && \left. \mbox{}
  + i x_1 \sigma_2 \frac{(x_1^2 + x_2^2 + L^2)}
  {8 L^2 \lambda^{\rm f}_{ 1}}
  \right. \\ && \left. \mbox{}
  - i x_2 \sigma_1 \frac{(x_1^2 + x_2^2 + L^2)}
  {8 L^2 \lambda^{\rm f}_{ 2}}
  \right] \nonumber .
\end{eqnarray}
With this transformation, the matrix vector $\mathbf a$ is mapped
to
\begin{eqnarray}
  \mathbf a &\to&  \mathbf{\hat e_1} \left[
  \frac{e B_3 x_2}{2 c} 
  + \frac{\hbar x_2 \sigma_3}{4 \lambda_1^{\rm c} \lambda_2^{\rm c}} 
  \right. \nonumber \\ && \left. \mbox{} 
  - \frac{\hbar}{8 L^2}
  \left(\frac{2 x_1 x_2 \sigma_1}{\lambda^{\rm f}_{ 2}} +
  \frac{(3 x_2^2 + x_1^2 - L^2) \sigma_2}{\lambda^{\rm f}_{ 1}}
  \right)
  \right]
  \nonumber \\
  && \mbox{} -  \mathbf{\hat e_2} \left[\frac{e B_3 x_1}{2 c} 
  + \frac{\hbar x_1 \sigma_3}{4 \lambda_1^{\rm c} \lambda_2^{\rm c}}  
  \right. \nonumber \\ && \left. \mbox{}
  - \frac{\hbar}{8 L^2}
  \left(\frac{2 x_1 x_2 \sigma_2}{\lambda^{\rm f}_{ 1}} +
  \frac{(3 x_1^2 + x_2^2 - L^2) \sigma_1}{\lambda^{\rm f}_{ 2}} \right)
  \right].
  \nonumber \\
  \label{eq:anonunif}
\end{eqnarray}
Comparing Eqs.\ (\ref{eq:anonunif}) and (\ref{eq:anew}), 
we immediately conclude that with a non-uniform 
spin-orbit scattering strength of the form (\ref{eq:nonunif})
with $\lambda^{\rm f} =
\lambda^{\rm f}_{ 1} = \lambda^{\rm f}_{ 2}$
one should make the identification
\begin{equation}
  a^2 =  
  \left( \frac{L^2}{4 (\lambda^{\rm f})^2} \right)
  \frac{v_F l \pi}{12 L^2 \Delta}.
\end{equation}
In terms of the energy scale $\varepsilon_{\Vert}^{\rm so}$, this
implies
\begin{equation}
  \varepsilon_{\Vert}^{\rm so}
  = \frac{1}{6} E_{\rm Th} \left( \frac{L^2}{4(\lambda^{\rm f})^2} 
\right),
\end{equation}
in agreement with the general relation (\ref{eq:kappageneral}).


\begin{thebibliography}{99}
\bibitem{RashbaDresselhaus} In conventional notation, the Hamiltonian is
specified in coordinates with respect to the $\hat x = [100]$ and $\hat y =
[010]$ directions. Then, the spin-orbit part of the Hamiltonian takes the
form 
$$
  H_{\mathrm{so}} = \frac{\gamma}{m}(p_x \sigma_y - p_y \sigma_x) + 
  \frac{\eta}{m} (p_x \sigma_x - p_y \sigma_y).
$$
Here $\gamma$ and $\eta$ are coupling constants for the Rashba and
Dresselhaus terms, respectively. In terms of the rates $\gamma$ and $\eta$,
one has $\hbar/2\lambda_1 = -\eta + \gamma$, $\hbar/2\lambda_2 = -\eta -
\gamma$.
\end{thebibliography}
\end{document}